\documentclass[twocolumn, astrosymb, resetfootnote]{aastex7}
\usepackage{xspace}
\usepackage{gensymb}
\usepackage{xcolor}
\usepackage{float}

\newcommand{\addcite}[1]{{\color{red}[CITE]}}
\newcommand{\pz}{photo-$z$\xspace}
\newcommand{\pzs}{photo-$z$'s\xspace}
\newcommand{\Pzs}{Photo-$z$'s\xspace}

\makeatletter
\newcommand\blfootnote[1]{%
  \begingroup
  \renewcommand{\thefootnote}{}
  \def\@makefntext##1{\noindent\normalfont##1}
  \footnote{#1}%
  \addtocounter{footnote}{-1}
  \endgroup
}
\makeatother

\shorttitle{DESI spectroscopy for photo-$z$ training \& calibration}
\shortauthors{B. Dey et al.}
\graphicspath{{./}{figures/}}
\suppressAffiliations
\begin{document}

\title{Deep Spectroscopy with DESI for Photometric Redshift Training and Calibration}

\correspondingauthor{Biprateep Dey}
\author[orcid=0000-0002-5665-7912,gname=Biprateep, sname=Dey]{Biprateep Dey}
\affiliation{Department of Statistical Sciences, University of Toronto, Toronto, ON M5G 1Z5, Canada}
\affiliation{Canadian Institute for Theoretical Astrophysics (CITA), University of Toronto, Toronto, ON M5S 3H8}
\affiliation{Dunlap Institute for Astronomy \& Astrophysics, University of Toronto, Toronto, ON M5S 3H4, Canada}
\affiliation{Department of Physics and Astronomy and PITT PACC, University of Pittsburgh, Pittsburgh, PA 15260, USA}
\email[show]{biprateep@pitt.edu}  

\author[orcid=0000-0001-8684-2222,gname=Jeffrey, sname=Newman]{Jeffrey A. Newman}
\affiliation{Department of Physics and Astronomy and PITT PACC, University of Pittsburgh, Pittsburgh, PA 15260, USA}
\email{janewman@pitt.edu}

\author[orcid=0000-0001-8684-2222]{Tianqing Zhang}
\affiliation{Department of Physics and Astronomy and PITT PACC, University of Pittsburgh, Pittsburgh, PA 15260, USA}
\email{tq.zhang@pitt.edu}

\author[0000-0000-0000-0000,gname='Jessica Nicole', sname='Aguilar']{J.~Aguilar}
\affiliation{Lawrence Berkeley National Laboratory, 1 Cyclotron Road, Berkeley, CA 94720, USA}
\email{jaguilar@lbl.gov}

\author[0000-0001-6098-7247, gname='Steven', sname='Ahlen']{S.~Ahlen}
\affiliation{Department of Physics, Boston University, 590 Commonwealth Avenue, Boston, MA 02215 USA}
\email{ahlen@bu.edu}

\author[0000-0003-2923-1585, gname='Abhijeet', sname='Anand']{A.~Anand}
\affiliation{Lawrence Berkeley National Laboratory, 1 Cyclotron Road, Berkeley, CA 94720, USA}
\email{abhijeetanand2011@gmail.com}

\author[0000-0000-0000-0000,gname='Brett H.', sname='Andrews']{B.~Andrews}
\affiliation{Department of Physics \& Astronomy and Pittsburgh Particle Physics, Astrophysics, and Cosmology Center (PITT PACC), University of Pittsburgh, 3941 O'Hara Street, Pittsburgh, PA 15260, USA}
\email{andrewsb@pitt.edu}

\author[0000-0003-4162-6619, gname='Stephen', sname='Bailey']{S.~Bailey}
\affiliation{Lawrence Berkeley National Laboratory, 1 Cyclotron Road, Berkeley, CA 94720, USA}
\email{stephenbailey@lbl.gov}

\author[0000-0001-9712-0006, gname='Davide', sname='Bianchi']{D.~Bianchi}
\affiliation{Dipartimento di Fisica ``Aldo Pontremoli'', Universit\`a degli Studi di Milano, Via Celoria 16, I-20133 Milano, Italy}
\affiliation{INAF-Osservatorio Astronomico di Brera, Via Brera 28, 20122 Milano, Italy}
\email{davide.bianchi1@unimi.it}

\author[0000-0000-0000-0000,gname='David', sname='Brooks']{D.~Brooks}
\affiliation{Department of Physics \& Astronomy, University College London, Gower Street, London, WC1E 6BT, UK}
\email{david.brooks@ucl.ac.uk}

\author[0000-0001-7316-4573, gname='Francisco Javier', sname='Castander']{F.~J.~Castander}
\affiliation{Institut d'Estudis Espacials de Catalunya (IEEC), c/ Esteve Terradas 1, Edifici RDIT, Campus PMT-UPC, 08860 Castelldefels, Spain}
\affiliation{Institute of Space Sciences, ICE-CSIC, Campus UAB, Carrer de Can Magrans s/n, 08913 Bellaterra, Barcelona, Spain}
\email{fjc@ice.csic.es}

\author[0000-0000-0000-0000,gname='Todd', sname='Claybaugh']{T.~Claybaugh}
\affiliation{Lawrence Berkeley National Laboratory, 1 Cyclotron Road, Berkeley, CA 94720, USA}
\email{tmclaybaugh@lbl.gov}

\author[0000-0002-2169-0595, gname='Andrei', sname='Cuceu']{A.~Cuceu}
\affiliation{Lawrence Berkeley National Laboratory, 1 Cyclotron Road, Berkeley, CA 94720, USA}
\email{acuceu@lbl.gov}

\author[0000-0002-0553-3805, gname='Kyle', sname='Dawson']{K.~S.~Dawson}
\affiliation{Department of Physics and Astronomy, The University of Utah, 115 South 1400 East, Salt Lake City, UT 84112, USA}
\email{kdawson@astro.utah.edu}

\author[0000-0002-1769-1640, gname='Axel ', sname='de la Macorra']{A.~de la Macorra}
\affiliation{Instituto de F\'{\i}sica, Universidad Nacional Aut\'{o}noma de M\'{e}xico,  Circuito de la Investigaci\'{o}n Cient\'{\i}fica, Ciudad Universitaria, Cd. de M\'{e}xico  C.~P.~04510,  M\'{e}xico}
\email{macorra@fisica.unam.mx}

\author[0000-0003-0928-2000, gname='John', sname='Della Costa']{J.~Della~Costa}
\affiliation{NSF NOIRLab, 950 N. Cherry Ave., Tucson, AZ 85719, USA}
\email{jmdc112596@gmail.com}

\author[0000-0002-4928-4003, gname='Arjun', sname='Dey']{Arjun~Dey}
\affiliation{NSF NOIRLab, 950 N. Cherry Ave., Tucson, AZ 85719, USA}
\email{arjun.dey@noirlab.edu}

\author[0000-0000-0000-0000,gname='Peter', sname='Doel']{P.~Doel}
\affiliation{Department of Physics \& Astronomy, University College London, Gower Street, London, WC1E 6BT, UK}
\email{apd@star.ucl.ac.uk}

\author[0000-0003-4992-7854, gname='Simone', sname='Ferraro']{S.~Ferraro}
\affiliation{Lawrence Berkeley National Laboratory, 1 Cyclotron Road, Berkeley, CA 94720, USA}
\affiliation{University of California, Berkeley, 110 Sproul Hall \#5800 Berkeley, CA 94720, USA}
\email{sferraro@lbl.gov}

\author[0000-0002-3033-7312, gname='Andreu', sname='Font-Ribera']{A.~Font-Ribera}
\affiliation{Instituci\'{o} Catalana de Recerca i Estudis Avan\c{c}ats, Passeig de Llu\'{\i}s Companys, 23, 08010 Barcelona, Spain}
\affiliation{Institut de F\'{i}sica d’Altes Energies (IFAE), The Barcelona Institute of Science and Technology, Edifici Cn, Campus UAB, 08193, Bellaterra (Barcelona), Spain}
\email{afont@ifae.es}

\author[0000-0001-9632-0815, gname='Enrique', sname='Gaztañaga']{E.~Gaztañaga}
\affiliation{Institut d'Estudis Espacials de Catalunya (IEEC), c/ Esteve Terradas 1, Edifici RDIT, Campus PMT-UPC, 08860 Castelldefels, Spain}
\affiliation{Institute of Cosmology and Gravitation, University of Portsmouth, Dennis Sciama Building, Portsmouth, PO1 3FX, UK}
\affiliation{Institute of Space Sciences, ICE-CSIC, Campus UAB, Carrer de Can Magrans s/n, 08913 Bellaterra, Barcelona, Spain}
\email{gaztanaga@gmail.com}

\author[0000-0003-3142-233X, gname='Satya ', sname='Gontcho A Gontcho']{Satya~{Gontcho A Gontcho}}
\affiliation{University of Virginia, Department of Astronomy, Charlottesville, VA 22904, USA}
\email{satya@virginia.edu}

\author[0000-0000-0000-0000,gname='Daniel', sname='Gruen']{D.~Gruen}
\affiliation{Excellence Cluster ORIGINS, Boltzmannstrasse 2, D-85748 Garching, Germany}
\affiliation{University Observatory, Faculty of Physics, Ludwig-Maximilians-Universit\"{a}t, Scheinerstr. 1, 81677 M\"{u}nchen, Germany}
\email{daniel.gruen@lmu.de}

\author[0000-0000-0000-0000,gname='Gaston', sname='Gutierrez']{G.~Gutierrez}
\affiliation{Fermi National Accelerator Laboratory, PO Box 500, Batavia, IL 60510, USA}
\email{gaston@fnal.gov}

\author[0000-0001-9822-6793, gname='Julien', sname='Guy']{J.~Guy}
\affiliation{Lawrence Berkeley National Laboratory, 1 Cyclotron Road, Berkeley, CA 94720, USA}
\email{jguy@lbl.gov}

\author[0000-0002-9136-9609, gname='Hiram K.', sname='Herrera-Alcantar']{H.~K.~Herrera-Alcantar}
\affiliation{Institut d'Astrophysique de Paris. 98 bis boulevard Arago. 75014 Paris, France}
\affiliation{IRFU, CEA, Universit\'{e} Paris-Saclay, F-91191 Gif-sur-Yvette, France}
\email{herreraa@iap.fr}

\author[0000-0002-6550-2023, gname='Klaus', sname='Honscheid']{K.~Honscheid}
\affiliation{Center for Cosmology and AstroParticle Physics, The Ohio State University, 191 West Woodruff Avenue, Columbus, OH 43210, USA}
\affiliation{Department of Physics, The Ohio State University, 191 West Woodruff Avenue, Columbus, OH 43210, USA}
\email{kh@physics.osu.edu}

\author[0000-0002-6024-466X, gname='Mustapha', sname='Ishak']{M.~Ishak}
\affiliation{Department of Physics, The University of Texas at Dallas, 800 W. Campbell Rd., Richardson, TX 75080, USA}
\email{mishak@utdallas.edu}

\author[0000-0003-0201-5241, gname='Dick', sname='Joyce']{R.~Joyce}
\affiliation{NSF NOIRLab, 950 N. Cherry Ave., Tucson, AZ 85719, USA}
\email{richard.joyce@noirlab.edu}

\author[0000-0000-0000-0000,gname='Robert', sname='Kehoe']{R.~Kehoe}
\affiliation{Department of Physics, Southern Methodist University, 3215 Daniel Avenue, Dallas, TX 75275, USA}
\email{kehoe@physics.smu.edu}

\author[0000-0002-8828-5463, gname='David', sname='Kirkby']{D.~Kirkby}
\affiliation{Department of Physics and Astronomy, University of California, Irvine, 92697, USA}
\email{dkirkby@uci.edu}

\author[0000-0003-3510-7134, gname='Theodore', sname='Kisner']{T.~Kisner}
\affiliation{Lawrence Berkeley National Laboratory, 1 Cyclotron Road, Berkeley, CA 94720, USA}
\email{tskisner@lbl.gov}

\author[0000-0001-6356-7424, gname='Anthony', sname='Kremin']{A.~Kremin}
\affiliation{Lawrence Berkeley National Laboratory, 1 Cyclotron Road, Berkeley, CA 94720, USA}
\email{akremin@lbl.gov}

\author[0000-0002-1134-9035, gname='Ofer', sname='Lahav']{O.~Lahav}
\affiliation{Department of Physics \& Astronomy, University College London, Gower Street, London, WC1E 6BT, UK}
\email{o.lahav@ucl.ac.uk}

\author[0000-0003-1838-8528, gname='Martin', sname='Landriau']{M.~Landriau}
\affiliation{Lawrence Berkeley National Laboratory, 1 Cyclotron Road, Berkeley, CA 94720, USA}
\email{mlandriau@lbl.gov}

\author[0000-0001-7178-8868, gname='Laurent', sname='Le Guillou']{L.~Le~Guillou}
\affiliation{Sorbonne Universit\'{e}, CNRS/IN2P3, Laboratoire de Physique Nucl\'{e}aire et de Hautes Energies (LPNHE), FR-75005 Paris, France}
\email{llg@lpnhe.in2p3.fr}

\author[0000-0002-3677-3617, gname='Alexie', sname='Leauthaud']{A.~Leauthaud}
\affiliation{Department of Astronomy and Astrophysics, UCO/Lick Observatory, University of California, 1156 High Street, Santa Cruz, CA 95064, USA}
\affiliation{Department of Astronomy and Astrophysics, University of California, Santa Cruz, 1156 High Street, Santa Cruz, CA 95065, USA}
\email{alexie@ucsc.edu}

\author[0000-0003-1887-1018, gname='Michael', sname='Levi']{M.~E.~Levi}
\affiliation{Lawrence Berkeley National Laboratory, 1 Cyclotron Road, Berkeley, CA 94720, USA}
\email{melevi@lbl.gov}

\author[0000-0003-4962-8934, gname='Marc', sname='Manera']{M.~Manera}
\affiliation{Departament de F\'{i}sica, Serra H\'{u}nter, Universitat Aut\`{o}noma de Barcelona, 08193 Bellaterra (Barcelona), Spain}
\affiliation{Institut de F\'{i}sica d’Altes Energies (IFAE), The Barcelona Institute of Science and Technology, Edifici Cn, Campus UAB, 08193, Bellaterra (Barcelona), Spain}
\email{mmanera@ifae.es}

\author[0000-0002-4279-4182, gname='Paul', sname='Martini']{P.~Martini}
\affiliation{Center for Cosmology and AstroParticle Physics, The Ohio State University, 191 West Woodruff Avenue, Columbus, OH 43210, USA}
\affiliation{Department of Astronomy, The Ohio State University, 4055 McPherson Laboratory, 140 W 18th Avenue, Columbus, OH 43210, USA}
\email{martini.10@osu.edu}

\author[0000-0002-4475-3456, gname='Jamie', sname='McCullough']{J.~McCullough}
\affiliation{Department of Astrophysical Sciences, Princeton University, Princeton NJ 08544, USA}
\email{jmccullough@princeton.edu}

\author[0000-0002-1125-7384, gname='Aaron', sname='Meisner']{A.~Meisner}
\affiliation{NSF NOIRLab, 950 N. Cherry Ave., Tucson, AZ 85719, USA}
\email{aaron.meisner@noirlab.edu}

\author[0000-0000-0000-0000,gname='Ramon', sname='Miquel']{R.~Miquel}
\affiliation{Instituci\'{o} Catalana de Recerca i Estudis Avan\c{c}ats, Passeig de Llu\'{\i}s Companys, 23, 08010 Barcelona, Spain}
\affiliation{Institut de F\'{i}sica d’Altes Energies (IFAE), The Barcelona Institute of Science and Technology, Edifici Cn, Campus UAB, 08193, Bellaterra (Barcelona), Spain}
\email{rmiquel@ifae.es}

\author[0000-0002-2733-4559, gname='John', sname='Moustakas']{J.~Moustakas}
\affiliation{Department of Physics and Astronomy, Siena University, 515 Loudon Road, Loudonville, NY 12211, USA}
\email{jmoustakas@siena.edu}

\author[0000-0000-0000-0000,gname='Adam', sname='Myers']{A.~D.~Myers}
\affiliation{Department of Physics \& Astronomy, University  of Wyoming, 1000 E. University, Dept.~3905, Laramie, WY 82071, USA}
\email{amyers14@uwyo.edu}

\author[0000-0000-0000-0000,gname='Justin', sname='Myles']{J.~Myles}
\affiliation{Department of Astrophysical Sciences, Princeton University, Princeton NJ 08544, USA}
\email{jmyles@princeton.edu}

\author[0000-0001-9070-3102, gname='Seshadri', sname='Nadathur']{S.~Nadathur}
\affiliation{Institute of Cosmology and Gravitation, University of Portsmouth, Dennis Sciama Building, Portsmouth, PO1 3FX, UK}
\email{seshadri.nadathur@port.ac.uk}

\author[0000-0003-3188-784X, gname='Nathalie', sname='Palanque-Delabrouille']{N.~Palanque-Delabrouille}
\affiliation{IRFU, CEA, Universit\'{e} Paris-Saclay, F-91191 Gif-sur-Yvette, France}
\affiliation{Lawrence Berkeley National Laboratory, 1 Cyclotron Road, Berkeley, CA 94720, USA}
\email{npalanque-delabrouille@lbl.gov}

\author[0000-0002-0644-5727, gname='Will', sname='Percival']{W.~J.~Percival}
\affiliation{Department of Physics and Astronomy, University of Waterloo, 200 University Ave W, Waterloo, ON N2L 3G1, Canada}
\affiliation{Perimeter Institute for Theoretical Physics, 31 Caroline St. North, Waterloo, ON N2L 2Y5, Canada}
\affiliation{Waterloo Centre for Astrophysics, University of Waterloo, 200 University Ave W, Waterloo, ON N2L 3G1, Canada}
\email{will.percival@uwaterloo.ca}

\author[0000-0001-7145-8674, gname='Francisco', sname='Prada']{F.~Prada}
\affiliation{Instituto de Astrof\'{i}sica de Andaluc\'{i}a (CSIC), Glorieta de la Astronom\'{i}a, s/n, E-18008 Granada, Spain}
\email{fprada@iaa.es}

\author[0000-0001-6979-0125, gname='Ignasi', sname='Pérez-Ràfols']{I.~P\'erez-R\`afols}
\affiliation{Departament de F\'isica, EEBE, Universitat Polit\`ecnica de Catalunya, c/Eduard Maristany 10, 08930 Barcelona, Spain}
\email{ignasi.perez.rafols@upc.edu}

\author[0000-0000-0000-0000,gname='Graziano', sname='Rossi']{G.~Rossi}
\affiliation{Department of Physics and Astronomy, Sejong University, 209 Neungdong-ro, Gwangjin-gu, Seoul 05006, Republic of Korea}
\email{graziano@sejong.ac.kr}

\author[0000-0002-1609-5687, gname='Lado', sname='Samushia']{L.~Samushia}
\affiliation{Abastumani Astrophysical Observatory, Tbilisi, GE-0179, Georgia}
\affiliation{Department of Physics, Kansas State University, 116 Cardwell Hall, Manhattan, KS 66506, USA}
\affiliation{Faculty of Natural Sciences and Medicine, Ilia State University, 0194 Tbilisi, Georgia}
\email{lado@phys.ksu.edu}

\author[0000-0002-9646-8198, gname='Eusebio', sname='Sanchez']{E.~Sanchez}
\affiliation{CIEMAT, Avenida Complutense 40, E-28040 Madrid, Spain}
\email{eusebio.sanchez@ciemat.es}

\author[0000-0000-0000-0000,gname='David', sname='Schlegel']{D.~Schlegel}
\affiliation{Lawrence Berkeley National Laboratory, 1 Cyclotron Road, Berkeley, CA 94720, USA}
\email{djschlegel@lbl.gov}

\author[0000-0000-0000-0000,gname='Michael', sname='Schubnell']{M.~Schubnell}
\affiliation{Department of Physics, University of Michigan, 450 Church Street, Ann Arbor, MI 48109, USA}
\affiliation{University of Michigan, 500 S. State Street, Ann Arbor, MI 48109, USA}
\email{schubnel@umich.edu}

\author[0000-0002-6588-3508, gname='Hee-Jong', sname='Seo']{H.~Seo}
\affiliation{Department of Physics \& Astronomy, Ohio University, 139 University Terrace, Athens, OH 45701, USA}
\email{seoh@ohio.edu}

\author[0000-0002-3461-0320, gname='Joseph Harry', sname='Silber']{J.~Silber}
\affiliation{Lawrence Berkeley National Laboratory, 1 Cyclotron Road, Berkeley, CA 94720, USA}
\email{jhsilber@lbl.gov}

\author[0000-0000-0000-0000,gname='David', sname='Sprayberry']{D.~Sprayberry}
\affiliation{NSF NOIRLab, 950 N. Cherry Ave., Tucson, AZ 85719, USA}
\email{david.sprayberry@noirlab.edu}

\author[0000-0003-1704-0781, gname='Gregory', sname='Tarlé']{G.~Tarl\'{e}}
\affiliation{University of Michigan, 500 S. State Street, Ann Arbor, MI 48109, USA}
\email{gtarle@umich.edu}

\author[0000-0000-0000-0000,gname='Benjamin Alan', sname='Weaver']{B.~A.~Weaver}
\affiliation{NSF NOIRLab, 950 N. Cherry Ave., Tucson, AZ 85719, USA}
\email{benjamin.weaver@noirlab.edu}

\author[0000-0001-9382-5199, gname='Noah', sname='Weaverdyck']{N.~Weaverdyck}
\affiliation{Lawrence Berkeley National Laboratory, 1 Cyclotron Road, Berkeley, CA 94720, USA}
\email{nweaverdyck@lbl.gov}

\author[0000-0003-2229-011X, gname='Risa', sname='Wechsler']{R.~H.~Wechsler}
\affiliation{Kavli Institute for Particle Astrophysics and Cosmology, Stanford University, Menlo Park, CA 94305, USA}
\affiliation{Physics Department, Stanford University, Stanford, CA 93405, USA}
\affiliation{SLAC National Accelerator Laboratory, 2575 Sand Hill Road, Menlo Park, CA 94025, USA}
\email{rwechsler@stanford.edu}

\author[0000-0001-5381-4372, gname='Rongpu', sname='Zhou']{R.~Zhou}
\affiliation{Lawrence Berkeley National Laboratory, 1 Cyclotron Road, Berkeley, CA 94720, USA}
\email{rongpuzhou@lbl.gov}

\author[0000-0002-6684-3997, gname='Hu', sname='Zou']{H.~Zou}
\affiliation{National Astronomical Observatories, Chinese Academy of Sciences, A20 Datun Road, Chaoyang District, Beijing, 100101, P.~R.~China}
\email{zouhu@nao.cas.cn}




\begin{abstract}
Deep spectroscopic samples can be used to improve photometric redshift (\pz) estimates and reduce uncertainties on redshift distributions. Such improvements can increase the cosmological constraining power of large imaging-based experiments such as the Vera C. Rubin Observatory's Legacy Survey of Space and Time (LSST) and mitigate what may be a limiting systematic effect. We present results from the ``DESI-Deep pilot'' program, which was designed to assess the capability of the Dark Energy Spectroscopic Instrument (DESI) on the 4m Mayall telescope to measure redshifts of galaxies as faint as expected lensing samples for early LSST data ($m_i \leq 24.5$). We find that DESI is remarkably efficient at this task, with redshift success rates comparable to the results of observations from 10m-class telescopes with only $\sim2\times$ longer integration time (rather than $\sim 8\times$ longer as would be expected from aperture-area scaling), while simultaneously achieving $\sim30$ times larger multiplexing. We also find that the signal-to-noise
ratio of the spectra scales as expected for background-limited observations even for the longest exposure times ($\sim 7$ hours) and faintest targets in the program. These results demonstrate that DESI could provide the definitive redshift sample for the early years of LSST with a modest investment of observing time. Based upon the results of this program, we provide updated predictions for the time required to collect benchmark samples for \pz training and calibration using a variety of spectroscopic facilities. Finally, we describe a potential ``DESI-Deep'' survey designed to train and calibrate \pzs for imaging experiments, and provide forecasts of its impact on cosmological inference. 
\end{abstract}




\section{Introduction} \label{sec:intro}
Photometric redshifts (\pzs) -- i.e., estimates of galaxy redshifts obtained from photometric data alone -- are essential to the success of ongoing and upcoming imaging-based astrophysical experiments. Despite the high precision of spectroscopic redshifts, acquiring spectra for billions of faint objects observed by modern imaging surveys is impractical. Therefore, projects like the Dark Energy Survey (DES; \citealt{DESSurvey2005}), the Hyper Suprime-Cam Subaru Strategic Program (HSC-SSP; \citealt{Aihara2018HSCSSPsurvey}), the Euclid Mission \citep{Laureijs2011EuclidSurvey}, the Nancy Grace Roman Space Telescope \citep{Akeson2019RomanSurvey}, and the Vera C. Rubin Observatory's Legacy Survey of Space and Time (LSST; \citealt{Ivezic2019LSST}) all rely heavily on photometric redshifts to achieve their scientific objectives. \Pzs enable such projects to construct three-dimensional maps of galaxy and matter distributions, study the formation and evolution of galaxy populations, and trace the Universe's expansion and structure growth over time. Improved \pzs with reduced scatter will ultimately lead to smaller random errors in the measurements of cosmological parameters from these experiments. \blfootnote{Author affiliations listed at the end of this article. Data \& code available here: \url{https://biprateep.github.io/desi-deep-pilot/} }However, if systematic errors in the \pz estimates are not controlled, they could overshadow all other sources of uncertainty in these experiments.

Beginning with \citet{Baum1957FirstPhotoz}, numerous techniques have been employed to estimate photometric redshifts. These include physics-based approaches that model galaxy spectral energy distributions (SEDs) against observed photometry, and empirical methods that establish a non-linear mathematical relationship between input photometry and redshift using a training set of galaxies with known $z$. For an overview of commonly used methods, the reader can refer to \citet{Salvato2019PhotozReview}, \citet{Brescia2021Photozreview} or \citet{NewmanGruen2022PhotozReview}.

Regardless of the method used, large sets of spectroscopic redshift measurements that are representative of the target galaxy population are needed to improve algorithm performance and enhance the characterization of $z$ distributions. For physics-based methods, these measurements are needed for refining galaxy SED templates and error models, devising priors, and correcting for telescope throughput and photometric zero-points, whereas empirical methods require a representative training set to work at all. Larger spectroscopic data sets can reduce random uncertainties in photometric redshift estimates for individual objects (i.e., improve \textit{training}) and also enable more stringent estimates of the redshift distributions of galaxies in tomographic bins (i.e., improve \textit{calibration}). The same data set can help improve on both fronts, provided that it is devoid of biases introduced by sample selection.

The need for such data sets was strongly emphasized in several submissions to recent community planning exercises. These included submissions to the Decadal Survey on Astronomy and Astrophysics 2020 (Astro2020) by \cite{Newman2019Astro2020} and \citet{Mandelbaum2019Astro2020} and to the US Community Study on the Future of Particle Physics (Snowmass 2021) by \citet{2022BlazekSnowmass}, \citet{2022SchlegelSnowmass} and \citet{SnowmassCF4}. This need was subsequently endorsed by the final report of the Particle Physics Project Prioritization Panel (P5; \citealt{P5Report}). The importance of spectroscopic samples for redshift calibration was further demonstrated in the recent Kilo-Degree Survey (KiDS; \citealt{Wright2024KIDSDR5}) Legacy analysis, which found that the incorporation of larger, higher-quality spectroscopic samples and improved methods of redshift calibration greatly reduced the level of tension with Planck cosmological results \citep{Wright2025KIDSLegacyCosmo,Wright2025KidsLegacyRedshift}.

Previous efforts, such as the Complete Calibration of the Color-Redshift Relation (C3R2; \citealt{Masters2017C3R2}) survey, have sought to optimize the targeting of spectroscopic samples by observing galaxies in specific regions of color space that are underrepresented or missing in existing data sets, thereby aiming to complete the map of the color-redshift relationship. Recent and proposed surveys such as the DESI Complete Calibration of the Color–Redshift Relation (DC3R2; \citealt{McCullough2024DC3R2}) and 4MOST Complete Calibration of the Color-Redshift
Relation (4C3R2; \citealt{Gruen20234C3R2}) also follow a similar approach. However, as the galaxy population evolves significantly over cosmic time, samples of brighter galaxies do not cover the full span of colors of fainter objects at the same $z$ (as demonstrated by \citet{Blanton2006GalaxyEvolution}). Furthermore, redshift can vary with magnitude for fixed galaxy colors \citep{McCullough2024DC3R2,Masters2017C3R2}. Therefore, we cannot simply use spectroscopic data sets that are limited to brighter galaxies for training and calibration of redshifts of fainter galaxies. Moreover, heavily color- (or color-magnitude-) selected samples, which often involve nonlinear selection criteria, can introduce selection biases that affect our ultimate redshift estimates in complex ways \citep{Gruen2017Biases,Hartley2020SpectroscopicIncompleteness,Myles2021DESY3Photoz}. Therefore, magnitude-limited spectroscopic samples that match the depths of the photometric samples used in a study and include minimal selection cuts can offer a way forward to mitigate these biases.

\citet{Newman2015SpectroscopicNeeds} laid out a proposal for such a sample arguing approximately 20,000 spectra, distributed over multiple wide sky fields and matching the depths of imaging surveys, should be sufficient to meet the \pz training and calibration requirements for next-generation astrophysical experiments like LSST; both past theoretical forecasts and the scaling of errors with training set size for machine learning algorithms lead to numbers of this order. They also provided estimates of the time required to obtain such samples calculated by appropriately scaling the exposure times required by the DEIMOS spectrograph \citep{Faber2003Deimos} on the Keck 10m telescope, while accounting for the multiplexing capabilities, fields of view, and collecting areas of various instruments. However, these scaling relations require testing now that a new generation of spectroscopic instruments, such as the Dark Energy Spectroscopic Instrument (DESI; \citealt{DESI2022Overview}) and the Subaru Prime Focus Spectrograph (PFS; \citealt{Tamura2016PFS}), are becoming available. This work aims to provide such an update, by leveraging the capabilities of DESI.

The Dark Energy Spectroscopic Instrument (DESI; \citealt{DESI2022Overview}) is a robotic, fiber-fed, highly multiplexed spectroscopic surveyor operating on the Mayall 4-meter telescope at Kitt Peak National Observatory. DESI can obtain spectra of nearly 5000 objects simultaneously over a $\sim3\degree$ diameter field \citep{DESI2016Overview-1,DESI2016Overview-2,Silber2023DESIFocalPlane,Miller2023DESICorrector,Poppett2024Fiber}. DESI is currently conducting an eight-year survey of roughly 17,000 sq. degs. of the sky which will obtain spectra for over 60 million galaxies and quasars (compared to the initial forecasts of $\sim$40 million). DESI is designed to enable a Stage IV dark energy experiment (by the definition of the Dark Energy Task Force;~\citealt{DETF2006}) with its main objective of revealing the nature of dark energy by achieving the most accurate measurement of the Universe's expansion history to date using baryonic acoustic oscillation measurements \citep{Levi2013DESISnowmass}. The First Data Release (DR1, \citealt{DESICollaborationDR1}) that includes spectra for more than 18 million unique targets is now public. Early results from DESI have already set significant constraints on models of dark energy \citep{DESIDR1Cosmo, DESIDR2Cosmo}.

Due to its high multiplexing capabilities, large field-of-view, and efficient design, DESI is well-positioned to deliver deep spectroscopic samples that could significantly enhance the scientific output from existing imaging-based dark energy experiments and early data from LSST. A secondary target program, the DESI Complete Calibration of the Color-Redshift Relation (DC3R2; \citealt{McCullough2024DC3R2}), has successfully utilized data from the survey to calibrate \pz distributions for color-selected and comparatively bright objects (with $z$-band magnitudes $\lesssim 22.1$). 

However, DESI's redshift measurement performance for fainter objects, particularly at LSST-like depths, is largely untested. This work addresses this need by studying DESI's performance for galaxies at depths comparable to those anticipated for the LSST Year 1 weak lensing sample ($m_{i} \lesssim 24.05$). We specifically targeted objects fainter than past DESI targets while utilizing observing strategies that help us achieve exposure times longer than in the DESI main survey. Our goal is to assess the feasibility of using the current DESI hardware (or improved versions thereof) to obtain spectroscopic samples that can effectively train photometric redshift methods and calibrate redshift distributions for current and future imaging experiments.

This article is organized as follows. In Section~\ref{sec:data}, we describe our target selection, observations, and data processing. In Section~\ref{sec:methods}, we detail our methods for estimating DESI's redshift measurement efficiency, scaling of the signal-to-noise ratio (SNR) with exposure time and source faintness, and creating a predictive model of the expected redshift success rate for a magnitude-limited sample. Section~\ref{sec:results} then presents the corresponding results. Section~\ref{sec:discussion} provides some intuition behind our results to aid in their interpretation and presents updated estimates of the time required to conduct a survey similar to that proposed by \citet{Newman2015SpectroscopicNeeds} using various instruments. Section~\ref{sec:discussion} also lays down an example plan for a future DESI survey to obtain such a sample and quantifies the benefits of such a data set in terms of cosmological constraining power. Finally, Section~\ref{sec:summary} summarizes the content of the article.

\section{Data} \label{sec:data}
The sample presented in this article, hereafter referred to as the ``DESI-Deep pilot sample'' (or simply the ``pilot sample'') was observed as part of a pilot program to characterize the potential to target new classes of objects with DESI after its current eight-year survey is complete. Two patches of the sky were chosen to carry out the observations, centered around the fields observed by the XMM-LSS \citep{XMMLSS} and COSMOS \citep{COSMOSSurvey} surveys because of the availability of additional multi-wavelength observational data in these regions. The field of view of DESI ($\sim$8 sq.\ degs.) is significantly larger than the portions of these fields with the most extensive imaging. As a result, many of the DESI fibers could only be devoted to target classes that could be selected with more limited photometry. This has enabled the tests of DESI's performance for broadly-selected samples of faint galaxies described in this paper.

For our purposes, we shall refer to the region between $33.5\degree \leq$ R.A. $\leq37.5\degree$ and $-7\degree\leq$ DEC. $\leq-3\degree$ as the DESI-XMMLSS field and the region between $148\degree \leq$ R.A. $\leq152\degree$; $0\degree\leq$ DEC. $\leq4\degree$ as the DESI-COSMOS field. Some additional observations were carried out in a region between $244\degree \leq$ R.A. $\leq248.5\degree$ and $41.5\degree\leq$ DEC. $\leq45\degree$ which we label the DESI-Hercules field. Those observations were carried out as a filler program because the spectrographs were not properly cooled and, therefore, yielded poor data quality. 
Results from this field nevertheless do not change the overall conclusions of this work; for completeness, we present selected analyses of the DESI-Hercules data in Appendix~\ref{app:hercules}. All results in the main body of this paper are based only on data from the DESI-XMMLSS and DESI-COSMOS fields. 

\subsection{Target Selection}
We have used photometry from the Hyper Suprime-Cam Subaru Strategic Program (HSC-SSP; \citealt{HSCSSP}) public data release 3 (PDR3; \citealt{HSCPDR3}) wide catalog to select targets for the observations. We first queried for all unique sources (i.e., those with \texttt{isprimary=true}) from the PDR3 catalog located in our regions of interest in the sky and implemented several quality cuts. Specifically, we require that objects have a positive and finite \texttt{cmodel} flux measurement in the $i$-band and that the $i$-band photometry model fitting process did not fail (i.e., that \texttt{i\_cmodel\_flag=false} and \texttt{i\_sdsscentroid\_flag=false}). We then correct all apparent magnitudes in this work for Milky Way dust extinction based upon the \citet{SFD1998DustMap} reddening maps. As a proxy for DESI $i$-band fiber magnitudes ($m_{i\text{-}\mathrm{fiber}}$), we use the quantity \texttt{i-convolvedflux-2-15-flux} from the HSC catalogs, which are the aperture magnitudes calculated from flux within a 1.5-arcsec (9.0-pixel) aperture diameter at a seeing of 1.09 arcsec (6.5 pixels FWHM). This definition differs slightly from the standard DESI convention for fiber magnitudes, which assumes the same aperture diameter but a fixed seeing of 1 arcsec. To assess the impact of this choice, we compared the $r$-band fiber magnitudes from HSC with those of the DESI Legacy Imaging Surveys DR9 \citep{SchlegelLSDR9} for objects within the DESI-COSMOS and DESI-XMMLSS footprints. We found a median offset of approximately 0.2 mag, with the HSC-based fiber magnitudes appearing systematically fainter. Consequently, the results presented in this work using HSC-based magnitudes should be considered conservative.

We then selected galaxies that have extinction-corrected $i$-band magnitudes in the ranges of $22\leq m_{i} \leq 24.5$ and $22 \leq m_{i\text{-}\mathrm{fiber}} \leq 24.75$ in the DESI-XMMLSS field, or $22\leq i \leq 24.5$ and $22 \leq m_{i\text{-}\mathrm{fiber}} \leq 25$ for objects in the DESI-COSMOS field. In order to ensure that the sample would not be dominated by faint objects but rather more evenly sample the underlying population, we randomly subsampled the selected galaxies to have a uniform distribution in $i$-band magnitude. This resulted in a list of targets with a density of 6,250 deg$^{-2}$ selected from the parent catalog with a density of approximately 97,500 deg$^{-2}$. We did not apply any strategy to photometrically separate stars from galaxies for these observations in order to avoid missing any galaxy/AGN populations. 

\subsection{Observations and Data Processing}
The observations were conducted using a series of individual exposures each lasting 1000 seconds of effective exposure time \citep{Schlafly2023SurveyOps, KirkbyETC}. This requires adapting the instrument exposure time depending upon the observing conditions to achieve the equivalent signal-to-noise ratio of what would have been obtained for observations with no moon, at the zenith, no Milky Way dust extinction, 1.1 arcsec seeing, and a sky background of 21.07 mag per square arcsec in the $r$ band in 1000 seconds. After each exposure, the telescope was dithered by 1.5 arcmin and a new set of fibers were allocated to the same targets. This was done to reduce the impact of any systematic errors in the data collection and processing pipeline. The list of potential targets was assigned specific fibers using the DESI fiber assignment algorithm \citep{RaichoorFiberassign}. To ensure that the fainter objects ($23\leq m_{i} \leq 24.5$) received longer total exposure times, such objects were assigned a higher priority within the DESI fiber assignment algorithm, though still lower than targets from other observational programs conducted simultaneously. As a consequence, more bright objects were observed in total, but their exposure time was shorter than that of objects in the fainter magnitude range.

The telescope dithering followed a strategy similar to the DESI survey validation (SV;~\citealt{DESI2024SurveyValidation}) in which the center of the focal plane was moved along a circle of radius 0.12\degree around the center of the field. The raw data were processed using the DESI spectroscopic pipeline \citep{Guy2023DESIPipeline}, and individual exposures for each object were co-added to obtain the final spectra. Subsequently, the redshifts were measured from each co-added spectrum using Redrock \citep{BaileyRedrock}. We used the DESI Data Release-1 (DR1; \citealt{DESICollaborationDR1}) software stack~\footnote{\url{https://data.desi.lbl.gov/doc/releases/dr1/software-version/}} for processing the DESI-Deep pilot data.

\begin{figure*}[h]
    \centering
    \includegraphics[width=\textwidth]{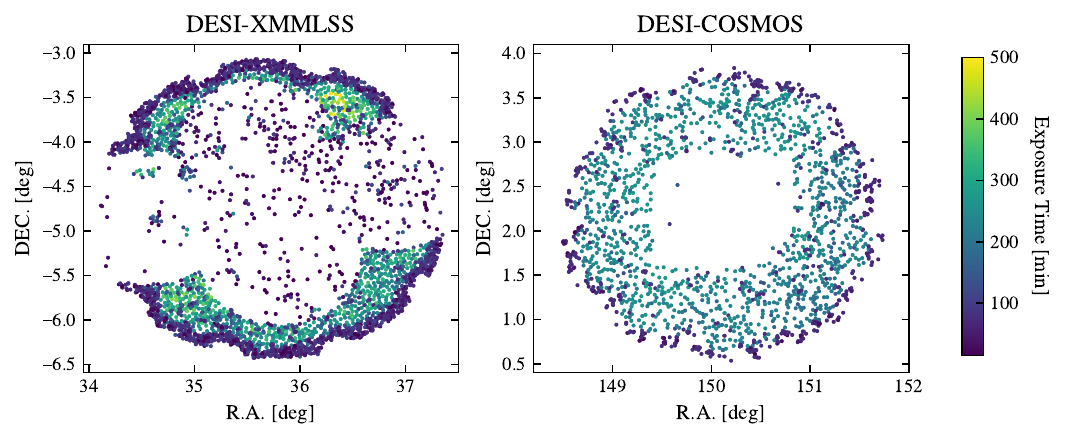}
    \caption{Locations of the 4863 objects observed in the DESI-XMMLSS ($33.5\degree \leq$R.A.$\leq37.5\degree$; $-7\degree\leq$DEC.$\leq-3\degree$) and DESI-COSMOS ($148\degree \leq$R.A.$\leq152\degree$; $0\degree\leq$DEC.$\leq4\degree$) fields. The colors of the points indicate the total effective exposure time for each object. The empty regions at the center of each field were filled by targets for other pilot observations for potential future DESI programs. }
    \label{fig:fibers_on_sky}
\end{figure*}

\begin{figure*}[h]
    \centering
    \includegraphics{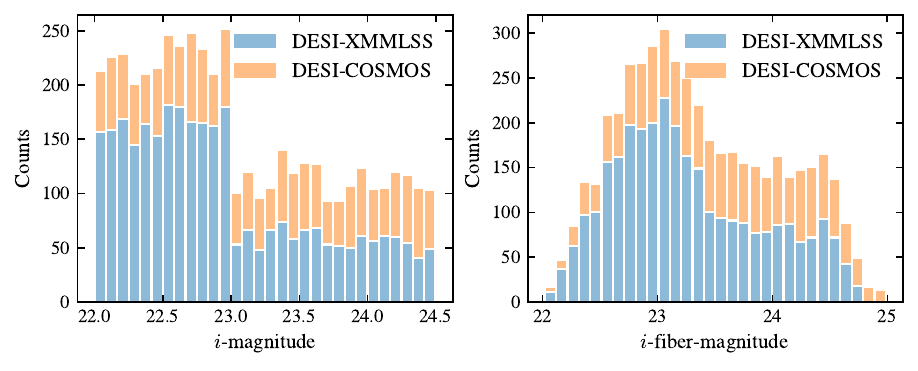}
    \caption{Distribution of $i$-magnitudes (left) and $i$-fiber-magnitudes (right) of the objects observed in the two fields, represented by stacked histograms. Objects with $i$-magnitude greater than 23 were allocated a higher priority in the DESI fiber assignment algorithm \citep{RaichoorFiberassign} resulting in them getting greater exposure times but for fewer objects.}
    \label{fig:mag-distribution}
\end{figure*}

\begin{figure}[h]
    \centering
    \includegraphics{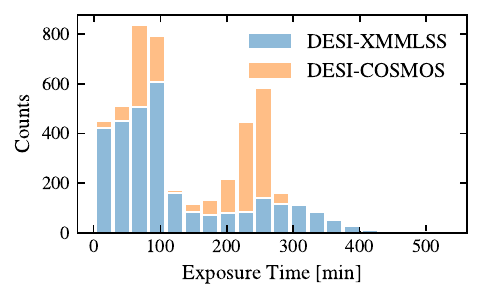}
    \caption{Distribution of effective exposure times represented by a stacked histogram. The median exposure time is about 94 min with the 5th and 95th percentiles of the distribution being 17 and 308 min respectively.}
    \label{fig:exp_time_dist}
\end{figure}

We observed a total of 4863 targets: 3095 in the DESI-XMMLSS field and 1768 in the DESI-COSMOS field. Observations for the DESI-XMMLSS field were conducted on October 29 and December 10, 14–16, 2022, while those for the DESI-COSMOS field occurred on March 16–17 and 24–26, 2023. Figure~\ref{fig:fibers_on_sky} shows the on-sky locations of the observed targets in the two fields along with their respective exposure times. Figure~\ref{fig:mag-distribution} shows the distribution of $i$-magnitude and $i$-fiber-magnitudes for the observed targets, and Figure~\ref{fig:exp_time_dist} shows the distribution of exposure times for the objects. The median effective exposure time was approximately 94 min (17–308 min), with a maximum of 535 min. The numbers in parentheses denote the 5th and 95th percentile values. For the DESI-XMMLSS observations, the median seeing was 1.04 arcsec (0.75–1.24 arcsec), and for the DESI-COSMOS observations, it was 1.09 arcsec (0.98–1.70 arcsec). The median airmass was 1.30 (1.25-1.45) for the DESI-XMMLSS observations and 1.24 (1.15-1.48) for the DESI-COSMOS observations. Similarly, the median transparency was 1.09 (0.92–1.14) for DESI-XMMLSS observations and 0.97 (0.90–1.02) for DESI-COSMOS observations.

\begin{figure*}
    \centering
    \includegraphics[]{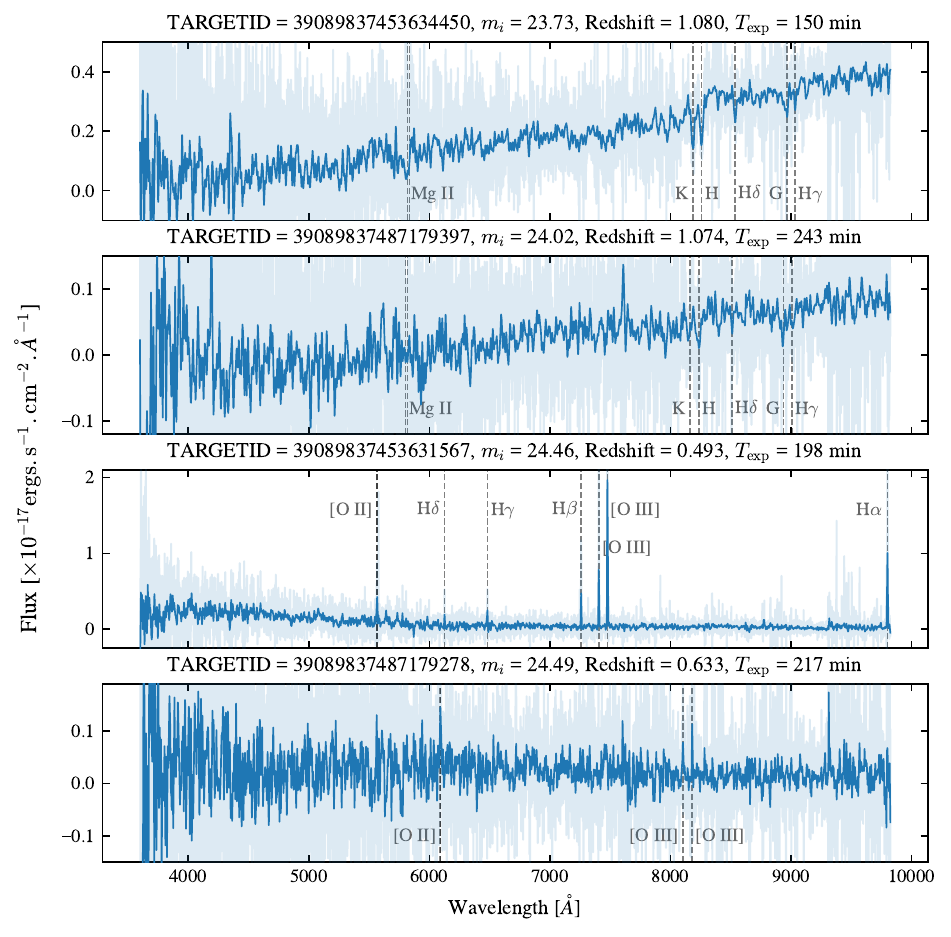}
    \caption{Spectra of four example galaxies from the DESI-Deep pilot sample, chosen to illustrate the diversity of galaxy types for which reliable redshifts were obtained. In each panel, the light blue shading represents the observed spectrum, while the solid dark blue line is the spectrum smoothed using an inverse-variance-weighted Gaussian kernel (with $\sigma=4${~\AA}  for the first two spectra and $\sigma=2.4${~\AA}  for the last two) to enhance spectral features for visualization. Data from the three DESI spectrograph arms have been coadded to form a single spectrum per object. Prominent absorption and emission lines are marked by labeled, gray dashed vertical lines. The TARGETID, $i$-band magnitude ($m_i$), measured redshift, and effective exposure time ($T_{\mathrm{exp}}$) are listed above each spectrum. The top two panels show spectra of quiescent galaxies with absorption features dominating the redshift determination, while the bottom two show star-forming galaxies with prominent emission lines. The first object shown from each class has a higher signal-to-noise ratio than the second. All of these objects are fainter than typical targets in the main DESI survey, demonstrating the instrument's ability to obtain reliable redshifts for galaxies as faint as those expected to be used for early lensing analyses with Rubin Observatory, given sufficient exposure time.
    }\label{fig:sample-spectra}
\end{figure*}

Figure~\ref{fig:sample-spectra} shows four example spectra from the DESI-Deep pilot sample which yielded reliable redshift measurements. The top two panels display spectra of quiescent galaxies (characterized by absorption features), while the bottom two panels show star-forming galaxies with prominent emission lines. Within each class, the first object shown has a higher SNR than the second, illustrating performance across a range of conditions. All of these objects are fainter than typical targets in the main DESI survey, demonstrating DESI’s capability to secure reliable redshifts for galaxies with magntitudes comparable to lensing samples expected from early LSST data. While obtaining reliable redshifts is typically easier for galaxies with strong emission lines, these examples show that with sufficient exposure time, DESI can also reliably detect the absorption features of faint quiescent galaxies. This capability is crucial, as it enables secure redshift measurements for galaxy populations that are frequently challenging to measure.

\section{Methods} \label{sec:methods}

The primary objective of this work is to evaluate DESI's effectiveness in measuring redshifts of objects substantially fainter than those typically observed during its main survey, reaching depths comparable to the faintest objects used for weak lensing analyses with LSST during its initial years of operation. We accomplish this by quantifying the redshift measurement efficiency of DESI as a function of both exposure time and target magnitude, while also assessing how the measured signal-to-noise ratios (SNRs) of the spectra scale with these parameters compared to the expectations for the background-limited regime. Additionally, we utilize the data to develop a model that establishes a relationship between exposure time, source brightness, and the probability of obtaining a successful redshift measurement, which could be used to plan future surveys with DESI or other similar instruments. 

\subsection{Estimating Redshift Measurement Efficiency}\label{sec:redshift-efficiency-method}

The DESI spectroscopic data reduction and redshift measurement pipelines, while extremely efficient, have only been optimized for the classes of objects observed during the main survey. Consequently, the standard automated metrics used to evaluate whether redshift measurements are secure were not well suited for the data obtained for this work. Instead, to determine whether the redshift for a given object was measured securely, we carried out a visual inspection (VI) campaign similar to what was done for DESI survey validation \citep{Lan2023GalaxyVI,Alexander2023QuasarVI}.

Specifically, the processed spectra, along with an initial redshift measurement from Redrock, were used to create interactive web pages using the Prospect~\footnote{\url{https://github.com/desihub/prospect}} spectrum visualization tool. Volunteers from the DESI collaboration then used the platform to visually inspect each spectrum with the Redrock fits and assigned a score of 0 to 4 to each spectra as a measure of the quality of the redshift measurement. A score of 4 indicated that two or more spectral features (e.g., emission or absorption lines or spectral breaks) are clearly and definitively visible, while 3 indicated the presence of at least one strong spectral feature as well as weaker spectral features that are highly likely to confirm the redshift. A score of 2 indicated that only one strong spectral feature was identifiable; 1 indicated that signal was present but the features could not be identified; and 0 indicated that no signal was present. If the automatic redshift measurement did not align with the observed spectrum, the inspectors manually determined the redshift by aligning a series of template spectra with the observed spectrum. 

\begin{figure}
    \centering
    \includegraphics{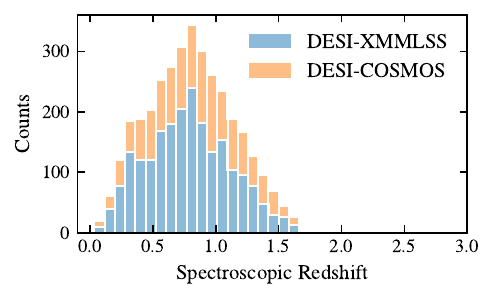}
    \caption{ The redshift distribution of objects for which spectra were observed and deemed good enough for reliable redshift measurements. Beyond redshift 1.6, most prominent spectral features of a galaxy shift out of the rest-frame observing range of the DESI spectrographs, leading to a lack of redshift measurements beyond this $z$. }
    \label{fig:redshift-distribution}
\end{figure}

Each spectrum was inspected by two volunteers and their scores were averaged. If the average score was either 3 or higher, or 2 or lower, the score was retained. For average scores between 2 and 3, a third inspection was conducted by an individual considered to be expert in redshift assessment, and their score was considered final. Since securely measuring a redshift requires two or more spectral features be identified, only those objects whose redshifts were assigned a final score of 3 or higher were considered good enough to make a successful redshift measurement. This binary division (into ``reliable'', i.e. quality $\geq 3$, versus ``unreliable'' redshifts) has been used to compute the instrument's redshift measurement success rates for the DESI-Deep pilot sample (see Sect.~\ref{sec:results-redshift-success}). To isolate the impact of target selection from instrument performance, we entirely excluded objects that had reliable redshift estimates but were spectroscopically classified as stars when calculating redshift measurement success rates. Figure~\ref{fig:redshift-distribution} illustrates the measured redshift distribution for objects deemed to have reliable redshift measurements after visual inspection. The redshift measurement success rate for some sample (e.g., objects within some bins in magnitude) is then simply the fraction of non-stellar objects in that sample that are assessed to have reliable redshift measurements.

\subsubsection{Adjusted Redshift Measurement Efficiency}\label{sec:adjustment-method}
Above $z\sim 1.6$, only spectral features that are weak in most galaxies will lie within the wavelength range covered by the DESI spectrographs.  As a result, objects may lack reliable redshift measurements due to their spectrum having insufficient signal-to-noise ratio to identify spectral features that are typically strong, or alternatively because they are intrinsically at a redshift greater than 1.6 where prominent features are not covered by the spectrum. To decouple these two effects, we therefore calculate an adjusted redshift measurement success rate by estimating the number of objects that would be expected to be at redshifts greater than 1.6 in a sample (e.g., a magnitude bin) and subtracting that from the number of failures and total number of objects. The adjusted success rate can be used to separately assess DESI's redshift measurement efficiency for a magnitude-limited sample with only objects at redshifts less than or equal to 1.6.

Since the redshift distribution of galaxies as a function of magnitude is not well measured at high redshifts, we have used two approaches to calculate the expected number of galaxies with redshift greater than 1.6 in a given bin, one analytic and one empirical. For an empirical adjustment, we used the COSMOS2020 data set \citep{Weaver2022Cosmos2020} with FARMER photometry \citep{Weaver2023Farmer} and restricted the data to only the ``clean'' objects (\texttt{FLAG\_COMBINED=0}). We used the redshift distribution of galaxies in this data set as a proxy for the true distribution. We used photometric redshift measurements obtained by the LePhare code \citep{Arnouts2002Lephare,Ilbert2006Lephare} and selected objects that have good photometric measurements and are consistent with being galaxies or active galactic nuclei.  The fraction of objects in a given magnitude bin that are at $z>1.6$ can then be calculated directly from these \pzs.

For the analytic adjustment, we instead use the empirical relations for galaxy redshift distributions provided in the LSST Science Book \citep{LSST2009LSSTScienceBook}, which were derived using the methods presented in \citet{Coil2004AngularClustering} (but based on the final DEEP2 dataset). In this case, the redshift probability density function for a  $i$-band magnitude limit $m_{i\text{-}\mathrm{lim}}$, $p(z|m_{i\text{-}\mathrm{lim}})$, is given by the formula:
\begin{equation}
    p(z|m_{i\text{-}\mathrm{lim}}) = \frac{1}{2z_{0}} \left(\frac{z}{z_{0}}\right)^{2}\exp[-(z/z_{0})^{\alpha}],
\end{equation}
where $z_{0} = 0.0417\times m_{i\text{-}\mathrm{lim}} - 0.744$ and $\alpha =1$. The total number of galaxies per square degree present within a magnitude-limited sample ($N_{\mathrm{gal}}$) is given by:
\begin{equation}
    N_{\mathrm{gal}}(m_{i\text{-}\mathrm{lim}}) = 46 \times 10^{0.31(m_{i\text{-}\mathrm{lim}}-25)}.
\end{equation}
By taking the integral of $p(z|m_{i\text{-}\mathrm{lim}})$ within specific magnitude bins and scaling that with $N_{gal}(m_{i\text{-}\mathrm{lim}})$, we can obtain the number counts of galaxies with redshifts greater than 1.6.

The analytical relations used are calibrated on optical spectroscopic data that become increasingly uncertain at $z > 1.5$. Similarly, photometric redshifts have not been fully validated for faint galaxy populations and  higher redshift regimes. Therefore, these adjusted success rates should be considered to be estimates based on currently available data which may have inaccuracies when extrapolated.

\subsection{Calculating Signal-to-Noise Ratios}\label{sec:snr-scaling-method}

In addition to assessing the redshift measurement efficiency, we also wish to investigate how the SNR of the observed spectra changes with exposure time and source brightness; this can then be compared to the expectations for observations in the background-limited regime (i.e., assuming that the noise in spectra is Poissonian and independent of source properties like flux, etc. but rather depends solely on the background count rate and exposure time). This analysis can provide insight into potential instrumental limitations, helping determine whether additional exposure time should yield meaningful improvements in SNR. 

To ensure the robustness of our findings, we performed an empirical measurement of the SNR which is independent of the DESI spectroscopic pipeline. The spectroscopic pipeline makes assumptions about the intrinsic spectra of galaxies and is optimized for the main survey targets \citep{Guy2023DESIPipeline}; we wish to avoid any impact from these assumptions. 

To calculate SNRs, we exclusively used data from the red arm of the DESI spectrographs in the wavelength range of 5900{~\AA}  to 7500{~\AA} . We focus on this wavelength range as it has been found in previous work \citep{Guy2023DESIPipeline} to be the most reliable regime for characterizing DESI instrument performance across a range of scenarios. We first subtract the best-fit Redrock template spectrum \citep{BaileyRedrock} from the spectrum of a given object. We then divide the wavelength range into 64 bins of 25{~\AA} width each and calculate the SNR for each bin as the ratio of the mean flux of the original spectrum within that bin divided by the standard deviation of the template-subtracted spectrum within it. The final empirical SNR is then calculated as the median of the binned SNRs. This yields a reliable quantification of the SNR of the spectral continuum, while being resilient to any anomalies, strong spectral features, etc. thanks to the robustness of the median statistic.

\subsection{Modeling the Redshift Success Rate}\label{sec:method-success-model}

Using the DESI-Deep pilot survey results, we have developed a simple model for the probability of obtaining a successful redshift measurement given the total DESI effective exposure time ($T_{\mathrm{exp}}$) and the brightness of a source. This model will be particularly valuable for planning future surveys with DESI or other similar instruments (cf. Sections~\ref{sec:updated-times} \& \ref{sec:future-survey}). To relate the exposure time and brightness to the success probability, we begin by defining a new parameter which we call the adjusted magnitude:
\begin{equation}
    \widetilde{m_i} = m_i - 1.25\log \left(\frac{T_{\mathrm{exp}}\mathrm{\  [sec]}}{6000\ \mathrm{[sec]}}\right),
\end{equation}
where $m_i$ is the extinction corrected $i$-band magnitude. This quantity is related to the negative logarithm of the SNR (Eq.~\ref{eq:snr-scale}) up to an additive constant.

The probability of success calculated from the outcome of a series of success–failure experiments (Bernoulli trials) follows a Binomial distribution. We therefore use a generalized linear model with a Binomial likelihood (also known as logistic regression; see~\citealt{McCullagh1989GLM} for a discussion on the topic) to model the observed success rate as a function of adjusted magnitude. For the systematic component of the model, we chose a linear transformation of $\widetilde{m_i}$, as it is the simplest model that well represents the data. The probability of obtaining a successful redshift measurement is then given by:
\begin{equation}
    p_{\mathrm{success}} = \sigma(\mathrm{A}.\widetilde{m_i}+\mathrm{B}),\label{eq:pr-success}
\end{equation}
where $\sigma(x)$ is the inverse of the logit function (also called the sigmoid function) and is defined as:
\begin{equation}
    \sigma(x) = \frac{1}{1+e^{-x}}
\end{equation}
and A and B are the parameters of the systematic component of the generalized linear model, which are determined using maximum likelihood estimation. We employ Bootstrap Resampling \citep{Efron1979Bootstrap} with 1000 resamples, for the construction of 95\% confidence and prediction intervals.

\section{Results}\label{sec:results}
 
\subsection{Redshift Success Rates}\label{sec:results-redshift-success}

\begin{figure}
    \centering
    \includegraphics{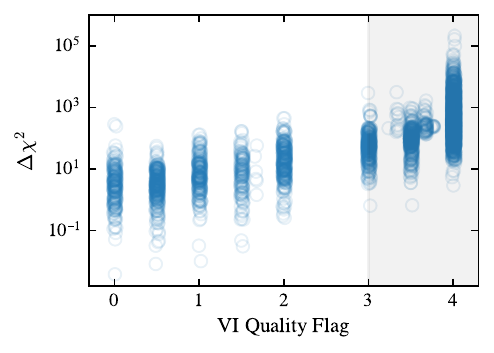}
    \caption{The $\Delta \chi^2$ between the best and second-best redshift fit plotted against visual inspection (VI) quality flags for our data set. The shaded gray region shows the flags corresponding to objects with reliable redshift measurements. While the DESI main survey uses $\Delta \chi^2$ for determining redshift measurement reliability, we observe a significant scatter between VI quality classifications and $\Delta \chi^2$ values, without any clear threshold that separates reliable redshift measurements (i.e., VI Quality Flag $\geq 3$) from unreliable ones for the DESI-Deep pilot data set. Therefore, we exclusively use VI labels for quality assessment in this work.}
    \label{fig:vi-vs-deltachi2}
\end{figure}

The DESI pipelines are optimized for the main survey targets, which are typically color-selected and brighter than the objects in the DESI-Deep pilot sample, potentially making the standard metrics used to assess the quality of a redshift measurement unreliable for our purposes. Although the main survey primarily employs $\Delta \chi^2$ thresholds \citep{DESI2024SurveyValidation} to assess redshift reliability (where $\Delta \chi^2$ represents the difference between the $\chi^2$ values for the best and second-best redshift fits), we observe from Figure~\ref{fig:vi-vs-deltachi2} that there is significant scatter in the relation between visual inspection quality flags and $\Delta \chi^2$ values; e.g., even some objects with $\Delta \chi^2 > 100$ (much larger than the threshold used for DESI samples) were found to have insecure redshift measurements, while many other objects with $\Delta \chi^2 < 10$ (below the typical DESI threshold) had reliable redshifts. 

Therefore, there is no clear threshold in $\Delta \chi^2$ that separates reliable redshifts from unreliable ones with high purity and completeness. Consequently, we rely on VI labels for quality assessment (as described in Section~\ref{sec:redshift-efficiency-method}) for this work. We are also working to develop automated methods for identifying reliable redshifts based on alternative metrics, which will be discussed in a future work.

We assess the instrument's redshift measurement performance using the redshift measurement success rate, defined as the fraction of objects that have reliable redshift estimates (i.e., VI quality flag $\geq 3$) and are spectroscopically classified as galaxies relative to the full number of non-stellar objects (i.e., excluding those with spectra reliably classified as stars from the denominator). This provides a conservative estimate of instrumental capability while isolating out any target selection effects. 

\begin{figure*}
    \centering
    \includegraphics{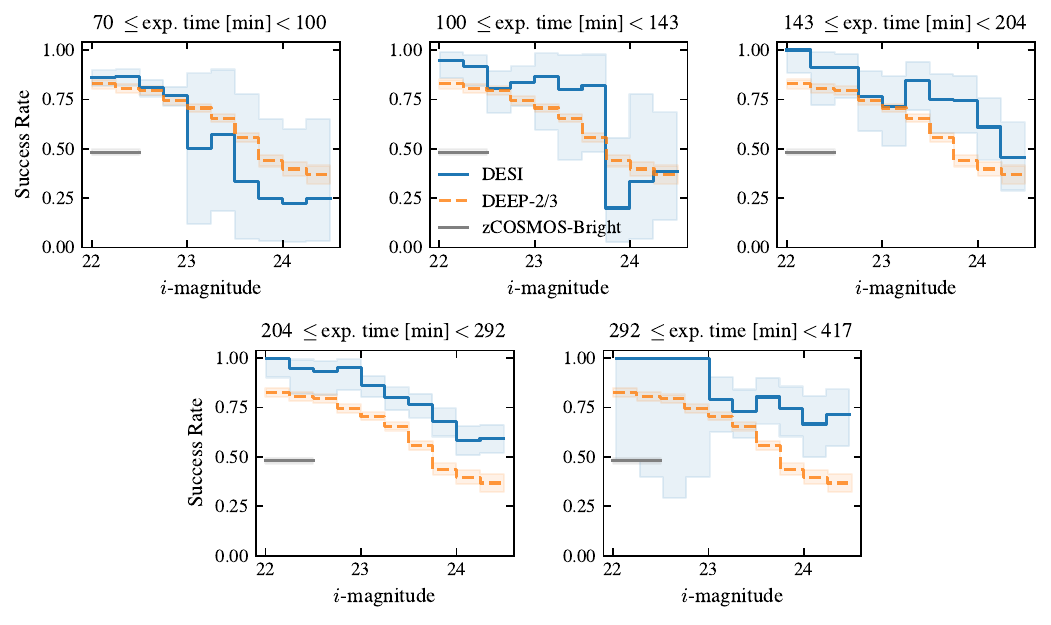}
    \caption{Redshift measurement success rate for the DESI-Deep pilot sample as a function of $i$-band magnitude, with separate panels corresponding to different ranges of effective exposure time. The blue curve shows the success rate obtained from our observations, with the 95\% confidence interval shown using the shaded blue region. The success rates as a function of $i$-magnitude for the combined DEEP2 and DEEP3 survey datasets are plotted in orange for comparison, with the shaded orange area showing the corresponding 95\% confidence interval. The gray line denotes the average success rate for the zCOSMOS-Bright sample for objects with $i$-band magnitudes between 22 and 22.5. DEEP2, DEEP3 and zCOSMOS-Bright had typical exposure times of roughly one hour, and hence the same curve is plotted for all bins of exposure time. We observe that at exposure times of about 2 hours DESI's success rate matches or exceeds that of the DEEP2 and DEEP3 galaxy redshift surveys. }
    \label{fig:i-mag-success}
\end{figure*}

\begin{figure*}
    \centering
    \includegraphics{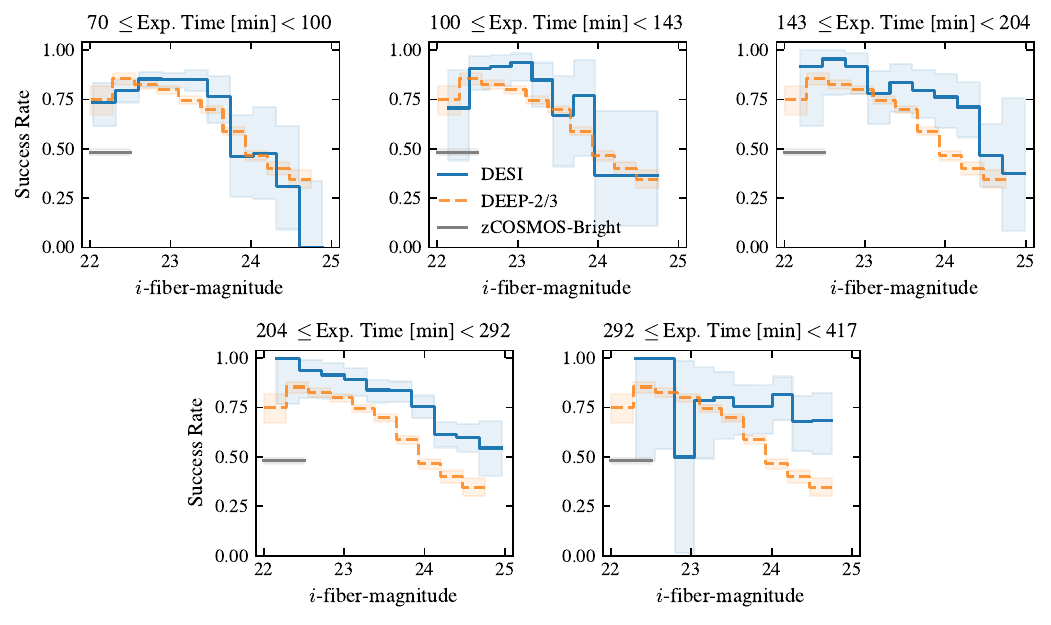}
    \caption{Redshift measurement success rate for the DESI-Deep pilot sample as a function of $i$-band fiber magnitude, with separate panels corresponding to different ranges of effective exposure time. The blue curve shows the success rate obtained for our observations, with the corresponding 95\% confidence interval shown as the shaded blue region. The success rates for the combined DEEP2 and DEEP3 survey dataset is plotted in orange for comparison, with the shaded orange area showing the corresponding 95\% confidence interval. The line denotes the average success rate for the zCOSMOS-Bright sample for objects with $i$-band magnitudes between 22 and 22.5. DEEP2, DEEP3 and zCOSMOS-Bright had typical exposure times of roughly one hour, and hence the same curve is plotted for all bins of exposure time. We observe that at exposure times of roughly 2 hours DESI's redshift measurement success rate matches that of DEEP2/DEEP3. }
    \label{fig:i-fiber-mag-success}
\end{figure*}

We have evaluated the redshift measurement success rate in bins of $i$-band magnitude, $i$-band fiber magnitude, or effective exposure time. Since these success rate estimates are derived from finite samples, we calculate the 95\% confidence limits using the Clopper-Pearson intervals \citep{ClopperPearson1934BinomialInterval}, which are appropriate for the binomial distribution that governs the probability of a particular spectrum yielding a reliable redshift. Figure~\ref{fig:i-mag-success} shows DESI's redshift measurement success rate as a function of $i$-band magnitude and in bins of effective exposure time. Figure~\ref{fig:i-fiber-mag-success} shows the same but as a function of the $i$-band fiber magnitude. Only observations with exposure times longer than one hour are included in each plot; the widths of the time bins were chosen such that the varying exposure times within a bin do not change the SNR by more than a factor of 1.2. 

We compare the DESI redshift measurement success rates with those of the DEEP2 and DEEP3 galaxy redshift surveys \citep{Newman2013Deep2,Zhou2019Deep3}. Data for these surveys were collected using the DEIMOS spectrograph \citep{Faber2003Deimos} on the Keck-II 10m telescope at W. M. Keck Observatory and represents the largest available (as of now) spectroscopic data set of galaxies at magnitudes similar to those of the DESI-Deep pilot sample.

We performed a spatial cross-match of the DEEP2/3 data with overlapping HSC-SSP PDR3 wide survey photometry with a tolerance of 1 arcsecond to perform the comparison. To create the parent catalog, we select objects with a photometric cross-match, spectroscopic quality flag (\texttt{ZQUALITY}) greater than 0, and measured redshifts greater than 0.001. We consider a DEEP2/3 redshift measurement to be successful if \texttt{ZQUALITY} is 3 or higher. The redshift measurement success rate is then defined as the fraction of objects with a successful redshift measurement in a given magnitude bin. All DEEP2/3 observations were performed with an on-sky exposure time of approximately 1 hour; therefore we do not bin this sample in time. 

We also compared the success rates with those from the zCOSMOS galaxy redshift survey \citep{Lilly2007zCOSMOS}; however, this test is limited by their sample that lacks a color selection (zCOSMOS-bright; \citealt{Lilly2009zCOSMOSbright}) having a magnitude limit ($m_{i}=22.5$) only a little fainter than the bright end of the DESI-Deep pilot sample.  However, where there is an overlap, the zCOSMOS-bright sample exhibits  significantly lower redshift measurement success rates (roughly 50\% for objects with $22<m_{i}\leq22.5$, when determined using similar criteria for a reliable measurement as this work) compared to either DEEP2/3 or this work at the same magnitudes.

Figures~\ref{fig:i-mag-success} and \ref{fig:i-fiber-mag-success} demonstrate that DESI consistently achieves a redshift measurement success rate greater than 75\% for brighter objects in our sample with relatively short exposure times of about 1.5 hours. Similar success rates are achieved for the fainter objects with high enough exposure times. Remarkably, we also see that with about 2 hours of exposure time, DESI can match the redshift measurement success rate of the DEEP2 and DEEP3 surveys. The observed trends are the same when the success rates are checked as a function of $i$-band magnitude or $i$-band fiber magnitudes, suggesting that the success rates are not significantly affected by the reduced amount of light entering the DESI fiber-fed spectrographs. 

\begin{figure*}
    \centering
    \includegraphics{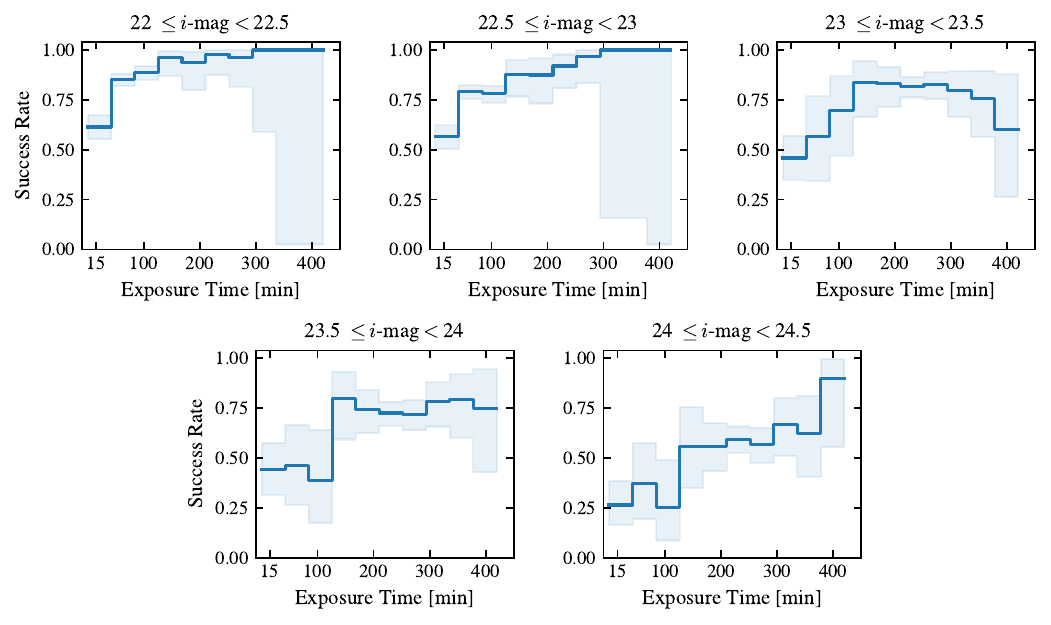}
    \caption{Redshift measurement success rate for the DESI-Deep pilot sample as a function of exposure time, with separate panels corresponding to different bins of $i$-band magnitude. The blue curve shows the success rate obtained for our observations, with 95\% confidence intervals shown using the shaded blue region. For every $i$ magnitude bin, we observe that the success rate increases with increase in exposure time as expected for a background-dominated measurement. The figure also provides us with a rough estimate of the  success rate to be expected for a given $i$-magnitude limit and exposure time.}
    \label{fig:exp-time-success}
\end{figure*}

Figure~\ref{fig:exp-time-success} shows the redshift measurement success rate as a function of exposure time in bins of $i$-band magnitude. As expected, we observe that the success rate increases with increasing exposure times at a given magnitude. We see that for the brighter objects in our sample over 75\% success rate can be achieved with about 2 hours of exposure time, whereas similar success rates require over 7 hours of exposure time for the faint objects in our sample. We have used our data to create a quantitative model of the success rate as discussed in Sections~\ref{sec:method-success-model} and~\ref{sec:success-model-results}.

\begin{figure*}
    \centering
    \includegraphics{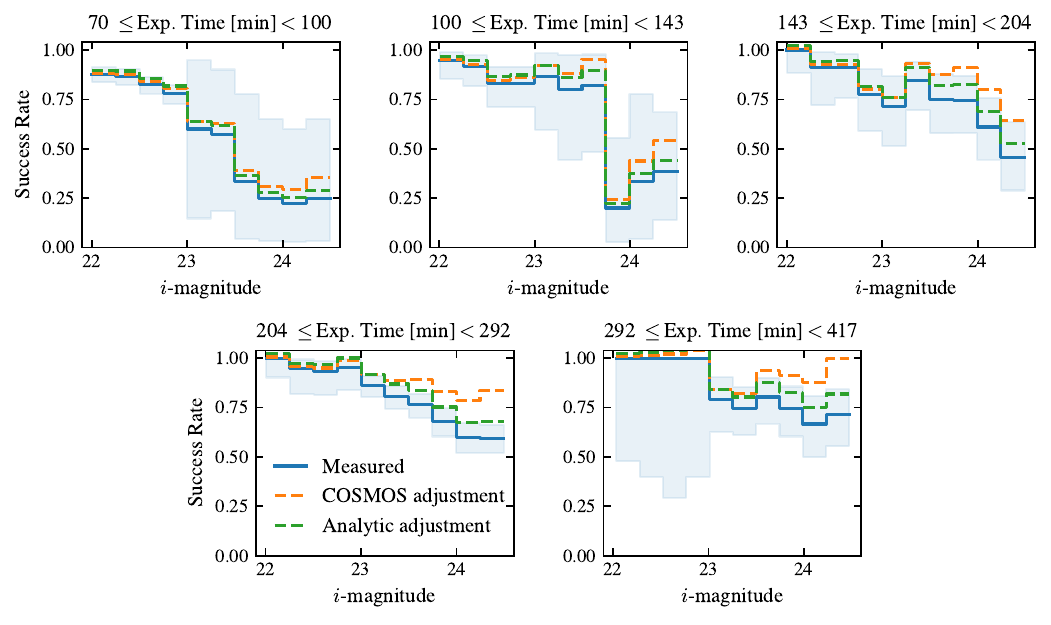}
    \caption{Test of the impact of adjusting for the abundance of $z>1.6$ objects when calculating redshift success rates.  The measured and adjusted redshift success rate are plotted as a function of $i$-band magnitude, with separate panels corresponding to different ranges of exposure time. As before, the solid blue curve shows the measured success rate as measured with 95\% confidence intervals shown using the shaded blue region. The dashed curves adjust this success rate to instead estimate the success rate for a magnitude-limited sample that only includes objects with redshifts less than or equal to 1.6. The orange dashed curve represents the success rate when the adjustment is calculated using the COSMOS2020 data set, while the dashed green line instead shows the success rate when the adjustment is done based upon analytic fits (see Section~\ref{sec:adjustment-method} for details). The adjusted curves should be indicative of DESI's redshift success rate when all the observed objects have prominent spectral features within the wavelength coverage of DESI. The effect of the $z>1.6$ failures becomes more prominent for the fainter objects in our sample, but even in that regime most redshift failures appear to be due to objects at redshifts below 1.6.}
    \label{fig:adjusted-success}
\end{figure*}

To estimate the redshift measurement success rate when the failures are only due to objects with redshift less than 1.6 -- i.e., objects with potential prominent spectral features within the rest-frame wavelength range accessible to DESI -- we can adjust the measured success rates following the procedure described in Section~\ref{sec:adjustment-method}. Figure~\ref{fig:adjusted-success} shows the measured and adjusted success rates as a function of $i$-band magnitude and in bins of exposure time with the adjustment done based upon data with deep multi-band photometry (COSMOS2020) or instead when based upon empirical analytic relations. The effect of the $z>1.6$ failures becomes prominent for the fainter objects in our sample, as the fraction of high-redshift objects increases in this regime. For the fainter objects in our data set, we find that roughly 5 to 10\% of the redshift measurement failures are due to $z>1.6$ objects, so the redshift failures are consistent with being dominated by objects with $z\leq1.6$.


\subsection{Scaling of Signal-to-Noise Ratios}

We next test how the scaling of noise in DESI spectra with exposure time compares to what would be expected if the data are background-limited.  To do this we can compare the observed scaling of the continuum SNR of the spectra (measured as described in Section \ref{sec:snr-scaling-method}) with exposure time and observed flux to what would be expected in the background-limited regime.  For that case, we expect:
\begin{displaymath}
    \mathrm{SNR} \propto \mathrm{Flux} \times \mathrm{(Exposure\ Time)}^{1/2}, 
\end{displaymath}
as the total number of signal counts will be proportional to flux times exposure time, whereas the noise is proportional to the square root of the number of background counts, which will be given by the background flux times the exposure time.  
Hence, for an object with $i$-band fiber magnitude $m_{i\text{-}\mathrm{fiber}}$ that is observed for an effective exposure time of $T_{\mathrm{exp}}$, the expected SNR can be written as:


\begin{equation}
    \mathrm{Expected\ SNR} = \kappa \times 10^{-(m_{i\text{-}\mathrm{fiber}}/2.5)}\times T_{\mathrm{exp}}^{1/2}, \label{eq:snr-scale}
\end{equation}
where $\kappa$ is a proportionality constant that we can determine from the data. In practice, we calculate its value separately for each object with an empirical SNR measurement, and then set $\kappa$ to be the median of these results. We use the fiber magnitude rather than overall magnitude of an object within Eq.~\ref{eq:snr-scale} as the former is more closely connected to the amount of light that enters the DESI spectrographs. We use Milky Way extinction-corrected fiber magnitudes for this calculation as the DESI dynamic exposure time calculator accounts for the effect of Galactic dust when scaling exposure times.

\begin{figure*}
    \centering
    \includegraphics{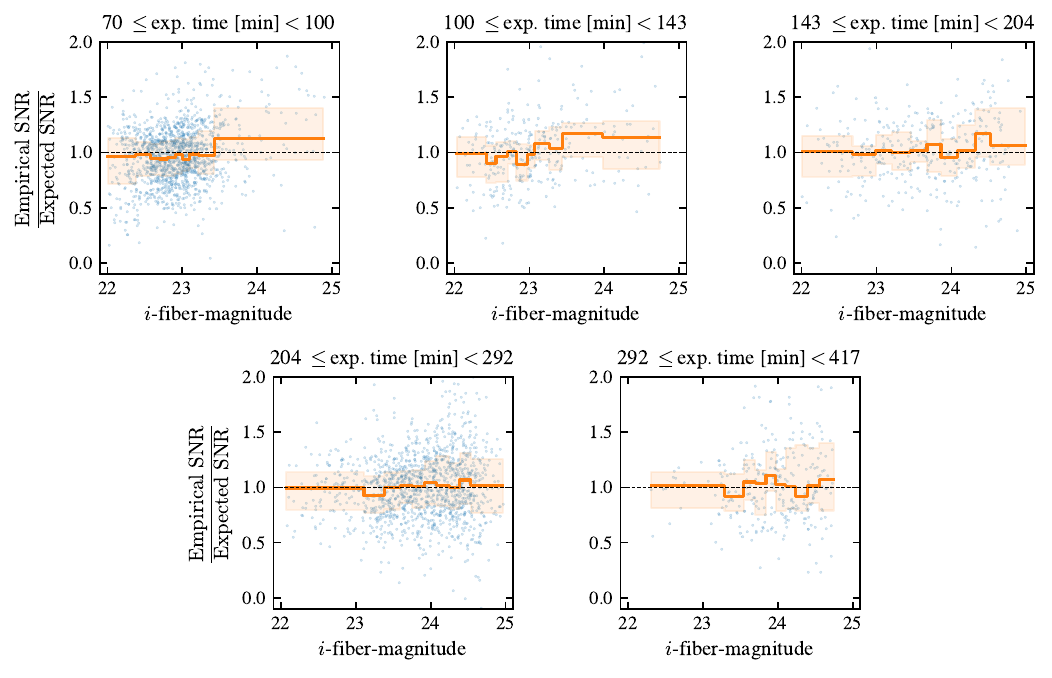}
    \caption{The ratio of empirically measured SNR and what we would expect from a background limited regime plotted as a function of $i$-fiber-magnitude in bins of exposure time. Each blue dot represents a single object and the orange curve denotes the average in ten equal population bins. The shaded orange region shows the 95\% confidence interval on the mean. We observe that the instrument performance follows the background limited expectations for the entire range of magnitude and exposure times tested and even the faintest galaxies and for the longest exposure times.}
    \label{fig:snr-v-mag}
\end{figure*}

\begin{figure*}
    \centering
    \includegraphics{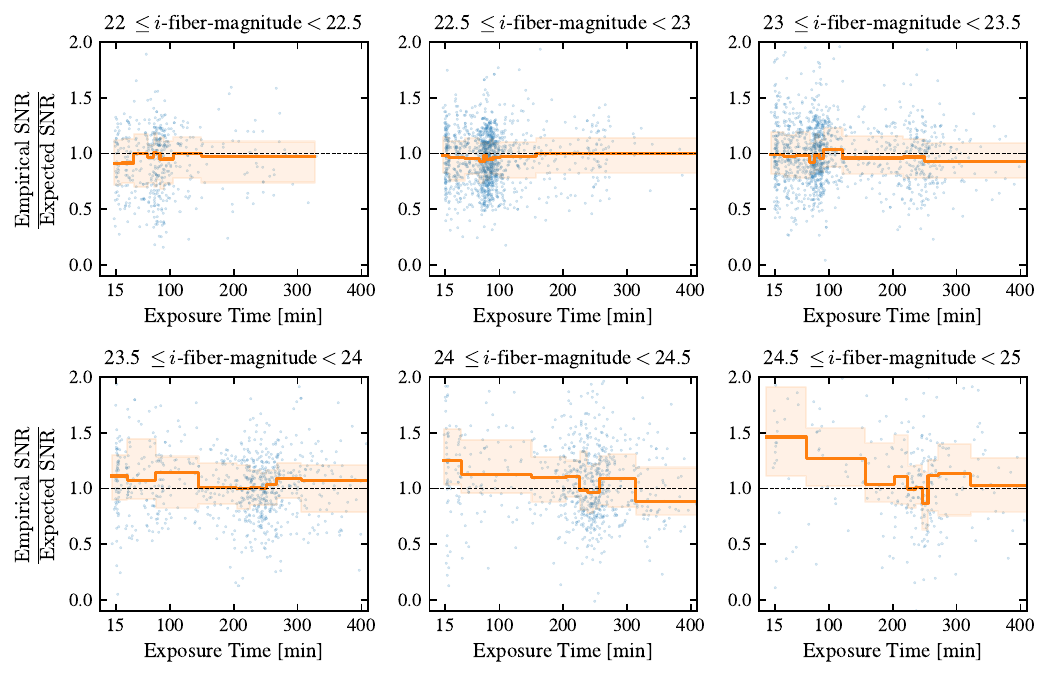}
    \caption{The ratio of empirically measured SNR and what we would expect from a background limited regime plotted as a function of exposure time in bins of $i$-fiber-magnitude. Each blue dot represents a single object and the orange curve denotes the average in ten equal population bins. The shaded orange region shows the 95\% confidence interval on the mean. We observe that the instrument performance follows the background limited expectations for the entire range of magnitude and exposure times tested and even the faintest galaxies and for the longest exposure times.}
    \label{fig:snr-v-time}
\end{figure*}

Figures~\ref{fig:snr-v-mag} and ~\ref{fig:snr-v-time} show that the ratio of empirical to expected SNR is close to one for the entire range of galaxy brightnesses and exposure times in our data set. DESI follows the expected SNR scaling of $T_{\mathrm{exp}}^{0.5}$, matching the scaling demonstrated using other similar fiber-fed spectroscopic instruments \citep[e.g.,][]{Lidman2020OzDES}. 
Achieving high redshift measurement success rates for target samples similar to lensing samples in LSST data will require even longer exposure times than those we have tested so far (cf. Section~\ref{sec:updated-times}).  However, the current results are very promising and indicate that it is likely that still deeper observations may be performed without immediately hitting a ceiling set by instrument systematics such as imperfect photon counting in the spectrograph CCDs, sky subtraction, or losses incurred due to the aperture of the fibers.



\subsection{Model for the Redshift Success Rate} \label{sec:success-model-results}

Following the methodology described in Section~\ref{sec:method-success-model}, we obtain the best-fit values and 95\% confidence intervals for the parameters in Equation~\ref{eq:pr-success} to be $\mathrm{A} = -1.20_{-0.11}^{+.10}$ and $\mathrm{B}= 28.74_{-2.28}^{+2.63}$. Therefore, our predictive model for quantifying the expected redshift success rate as a function of source magnitude and exposure time is given by:

\begin{equation}
    p_{\mathrm{success}} = \sigma(-1.20~ \widetilde{m_i}+28.74) \label{eq:pr-success-fit}
\end{equation}
 where $p_{\rm success}$ is the probability a particular object's redshift will be securely measured, $\widetilde{m_i} = m_i - 1.25\log \left(\frac{T_{\mathrm{exp}}\mathrm{[sec]}}{6000\ \mathrm{[sec]}}\right)$ and $\sigma$ is the sigmoid function defined as $\sigma(x) = \frac{1}{1+e^{-x}}$.

 For planning future surveys, it can be useful to invert this equation and reframe the same model to predict the required exposure time ($\widetilde{T}_{\mathrm{exp}}$) to achieve a success rate of $p_{\mathrm{success}}$ for an $i$-band magnitude limit of $m_{i}$. This is given by:
\begin{equation}
    \widetilde{T}_{\mathrm{exp}}(m_{i},p_{\rm success})\ \mathrm{[sec]} = 10^{-15.38+0.8~m_{i}+0.67\times\mathrm{logit}(p_{\rm success})} \label{eq:desi-exp} 
\end{equation} 
where the logit function is the inverse of the sigmoid function, defined as $\mathrm{logit}(x)=\ln(\frac{x}{1-x})$.

\begin{figure}[!h]
    \centering
    \includegraphics{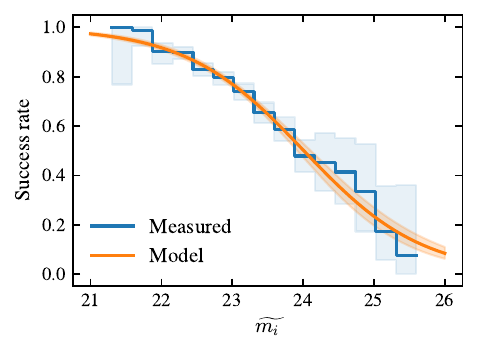}
    \caption{Redshift success rate versus the adjusted magnitude ($\widetilde{m_i} = m_i - 1.25\log \left(\frac{T_{\mathrm{exp}}\mathrm{\  [sec]}}{6000\ \mathrm{[sec]}}\right)$). The blue curve shows the success rate in 15 equal width bins, with the blue shaded region representing the $95\%$ confidence interval on the counts. The orange curve shows the generalized linear model fit to the data with the shaded orange region denoting the $95\%$ prediction interval obtained from bootstrap resampling. We have chosen the simplest model that well represents the observations.}
    \label{fig:success-v-adjusted-mag}
\end{figure}

Figure~\ref{fig:success-v-adjusted-mag} shows our best-fit predictive model along with the measured success rates in bins of effective magnitude. We observe that this simple model is able to represent the observed success rates very well and will allow us to make estimates of exposure times required for any future sky survey with a DESI-like instrument.

\section{Discussion}\label{sec:discussion}
Our results have shown that DESI has achieved a high redshift measurement efficiency and excellent signal-to-noise scaling compared to prior instruments. In this section, we provide insight into these results by discussing key differences in instrumentation and data processing pipelines that may explain these differences. We then use our measurements to update the survey time estimates for obtaining the redshift training and calibration samples proposed by \citet{Newman2015SpectroscopicNeeds} and \citet{2022BlazekSnowmass}. Finally, building on these findings, we lay out a plan for a new magnitude-limited survey, which we call the ``DESI-Deep'' program, to leverage the instrument's powerful capabilities for this purpose.

\subsection{Explaining Differences in Redshift Measurement Efficiency}
The results from the DESI-Deep pilot sample indicate that DESI must be significantly more efficient than the Keck/DEIMOS-based DEEP2 and DEEP3 surveys at measuring redshifts for faint galaxies, when accounting for differences in telescope aperture. The same holds for the VLT/VIMOS-based zCOSMOS survey, but in this section we will focus on DEEP2/DEEP3 where there is greater overlap in magnitude range and more detailed comparisons are feasible. This significant gap warrants investigation into its underlying causes. While a detailed, component-by-component comparison of the two facilities is complex and beyond the scope of this work, we discuss a few key factors in instrumentation, data processing and observing strategy that likely contribute to these differences, based on our experience with both DESI and DEIMOS.

One of the primary factors determining an instrument's efficiency for measuring redshifts is its total throughput, defined as the ratio of the number of photons captured by the spectrograph's detectors to those collected by the telescope's primary mirror. The Keck/DEIMOS system used in the DEEP2/3 surveys had a throughput of 15--25\%~\footnote{\url{https://www2.keck.hawaii.edu/inst/deimos/ripisc.html}}. In contrast, the DESI instrument achieves a total throughput of 30--40\% \citep{DESI2022Overview}, a notable figure given that fiber-fed spectrographs typically have had lower throughput compared to slit-based spectrographs. This enhanced throughput results from improvements across nearly all aspects of instrumentation, including advancements in corrector lens materials and designs, improvements in spectrograph design and stability, and the use of charge-coupled device (CCD) detectors with higher quantum efficiency compared to previous generations.

The effective throughput of spectroscopy will also be affected by any offsets between an object's true position on the sky and its targeted location, which would cause a significant fraction of light to fail to be captured by a fiber or slitlet. The imaging catalogs used for targeting by the DEEP2/3 surveys employed astrometry tied to the USNO-A2.0 catalog \citep{Monet1998USNOA2}; due to the sparseness of USNO-A2.0 stars, systematic errors in that catalog, and imperfections in the algorithms used to map this astrometry onto CFHT-12K imaging, those catalogs exhibited systematic astrometric errors of up to 1~arcsec, modulated over scales of 5--10~arcmin, when compared to SDSS catalogs \citep{Coil2004AngularClustering,Matthews2013DEEP2Photometry}. Since the slitmasks used by Keck/DEIMOS are 16~arcmin long and astrometric reference stars were near the edges of the masks, this could cause the masks to be misaligned compared to some objects.  Stars were typically centered within their alignment boxes to $\sim 0.1$~arcsec, but doing so could cause galaxies to be off center if they were in a different area of the imaging with relative astrometric offsets. 

In contrast the HSC-SSP PDR3 catalog we use for targeting has been able to base its global astrometry on the Gaia DR1 catalog \citep{GaiaCollaboration2016GaiaDR1}, which contributes negligibly ($\sim$1~milliarcsec) to the total error budget; the RMS scatter of bright star positions in HSC-SSP PDR3 relative to Gaia is 0.013~arcsec \citep{HSCPDR3}. The RMS error in positioning DESI fibers on the sky is 0.19~arcsec RMS \citep{Silber2023DESIFocalPlane} and contributes the most to the astrometric error budget. This large difference in astrometric accuracy could lead to substantial light losses with the narrow slits used for DEEP2/DEEP3 observations compared to DESI.

DESI adjusts the actual wall-clock exposure time for each observation dynamically, based on real-time observing conditions and Milky Way dust extinction, in order to achieve a relatively uniform SNR across exposures \citep{Schlafly2023SurveyOps}. As a consequence, the actual exposure times are generally somewhat longer than the effective exposure time values. In contrast, the DEEP2/3 surveys did not employ a dynamic exposure time calculator, although exposure times were occasionally extended on an ad hoc basis to compensate for degraded observing conditions. A typical DEEP2/3 exposure time was approximately 1 hour. For the DESI-Deep pilot sample, the median wall-clock time for observations was approximately 1.5 times larger than the corresponding effective exposure time. These differing strategies for determining exposure times may partly explain the observed differences in redshift measurement efficiency. However, even under the most favorable assumptions, a factor of 1.5 in effective exposure time is insufficient to fully explain the total difference in redshift measurement efficiency between the surveys.

Higher redshift measurement success rates can also be facilitated by better data processing, particularly the subtraction of night sky lines. DESI has significant instrumental advantages for this over DEIMOS, since the fibers are always the same (as opposed to the custom slitmasks used for each DEIMOS observation) and the spectrographs are kept stable (rather than having to rotate as the telescope moves in the case of DEIMOS).  The DESI spectroscopic data reduction pipelines also incorporate significant technical advances \citep{Guy2023DESIPipeline} which should contribute to its high redshift measurement efficiency. Two key improvements are the methods used for spectral extraction and sky subtraction.

The DEEP2/3 pipelines used the ``optimal extraction'' method from \citet{Horne1986OptimalExtraction}, which is based upon the assumption that the 2D spectrograph point-spread function (PSF) is separable, a condition often not met in practice. In contrast, DESI employs the ``spectro-perfectionism'' algorithm of \citet{Bolton2010Spectroperfectionism}. This technique performs a ``perfect'' extraction of one-dimensional spectra by creating a 2D forward model of the raw pixel data, using a precise model of the 2D PSF instead of a projected 1D profile. This approach yields several advantages, including minimizing variance, incorporating a common wavelength grid for all fibers, and having a well-defined resolution matrix. These improvements are crucial for subsequent processing steps, leading to better sky subtraction, flux and wavelength calibration, and ultimately, redshift estimation.

These advances in spectral extraction directly enable a more sophisticated approach to sky subtraction. The DEEP2/3 pipelines primarily utilized a ``local'' sky subtraction, where an interpolated signal from sky regions adjacent to the science target on a given slitlet was subtracted. By leveraging the common wavelength grid and resolution matrix from spectral extractions, the DESI pipeline utilizes ``non-local'' sky subtraction methods instead. It constructs a coherent model of the sky spectrum for an entire spectrograph image using data from numerous sky fibers distributed across the focal plane. The created model is further adjusted using a principal component analysis (PCA) derived model of strong sky lines (c.f. section 4.7 of \citealt{Guy2023DESIPipeline} for details). This comprehensive model is then subtracted from each science spectrum. The combination of ``spectro-perfectionism'' and non-local sky modeling allows the DESI pipeline to perform significantly better sky subtraction. Similar algorithms that utilize non-local sky subtraction with PCA-based methods have been shown to be effective in other fiber-fed spectroscopic instruments, enabling long exposure times without a significant reduction in expected SNR \citep{Sharp2010SkySubtractionDithering,Hart2019SkyResiduals}. Residuals from imperfect sky subtraction are the dominant source of systematic errors for ground-based, fiber-fed spectrographs that prevent efficient observations of faint sources \citep{Bundy2022FiberSpectraStability}. Reducing this primary source of errors helps to enable DESI to carry out deeper observations while still tracking the expected SNR scaling relations for background-limited spectroscopy.  

In addition to improvements in throughput and data processing, DESI's broad wavelength coverage significantly enhances its redshift measurement efficiency compared to previous surveys. The DESI spectrographs cover a range of nearly $6200${~\AA} (from $3600${~\AA} to $9800${~\AA}); in contrast, the DEEP2 survey with DEIMOS was limited to a span of about $2600${~\AA}. While the subsequent DEEP3 survey utilized a lower-resolution grating yielding a broader wavelength range, its effective coverage of approximately $4000-4500${~\AA} remains considerably smaller than that of DESI. This limited wavelength range restricts the spectral features observable at a given redshift, thereby lowering the overall redshift success rate. For instance, even if a galaxy has strong spectral features, those features may be redshifted out of the instrument's observable window. For the DEIMOS setups used in DEEP2/3, there was a practical maximum redshift limit of $z \approx 1.45$ compared to $z \approx 1.6$ for DESI, contributing to DESI's higher redshift measurement efficiency compared to the DEEP2/3 surveys.

Furthermore, DESI is able to provide this broader wavelength coverage while still maintaining a relatively high spectral resolution, particularly for the infrared arm of the spectrograph (covering $6560 - 9800${~\AA} with $R \sim 4000$). DESI can therefore resolve narrow features such as the [\ion{O}{2}] $\lambda\lambda3726,3729$ doublet, which is a key feature for redshift determination for galaxies with emission lines, at $z>0.76$.  In contrast, for the lower-resolution grating used in DEEP3, this feature is often unresolved. The contrast is even greater with lower-resolution instruments like the VIMOS spectrograph \citep{Lefevre2003VIMOS}; e.g., zCOSMOS-bright observations had resolution $R \sim 600$, much too low to split the two components of the [O II] doublet \citep{Comparat2013OIIDoublet}. Since any pair of spectral features (including both components of a single doublet) is sufficient to securely determine a redshift, this greatly benefits DESI's efficiency at delivering robust $z$ measurements.  The relative impact of resolution versus wavelength range is not entirely clear, however; both DEEP2 and DEEP3 delivered similar overall redshift measurement success rates to each other, despite differing by roughly a factor of two in spectral resolution and wavelength range.

A final source of differences between the DESI and DEEP2/3 observations lies in the observing strategies used.   For DEEP2/3 the same slitmask was typically used for all observations of a given object.  Furthermore, the instrument's flexure compensation system is designed to ensure that observations of an object are repeatably placed at the same position on the detector.  This can cause persistent patterns in sky subtraction residuals across observations (see Fig. 26 of \citealt{Newman2013Deep2} for examples).

In contrast, our pilot survey implemented an observing strategy designed to mitigate systematics in deep integrations. Total integrations, reaching up to a maximum of 7 hours, were broken into units of 1000-sec effective exposure time (the DESI exposure time calculator [ETC; \citealt{Schlafly2023SurveyOps,KirkbyETC}] is used to actively adjust exposure times to account for real-time observing conditions, including seeing, sky transparency, airmass, and Galactic reddening, keeping the depth equivalent between observations). Between each exposure, the telescope was slightly dithered, which places the same target onto a different fiber for subsequent observations. Thanks to DESI's fast readout and fiber positioning speeds, this strategy adds negligible overhead to the total observing time. The crucial benefit of this dithering approach is that it averages over individual fiber- and position-dependent instrumental systematics, including any residuals from sky subtraction, suppressing their impact on the final coadded spectrum by a factor of  (Number of exposures)$^{-\frac{1}{2}}$. We believe that this multi-exposure dithering strategy is a primary reason why the signal-to-noise ratio in our data scales as expected for purely background-limited observations, even for the longest total exposure times on the faintest objects.

It is uncertain which if any of these factors is the dominant explanation for why the redshift efficiency of DESI is so high compared to what would be expected from scaling prior Keck and VLT surveys by relative aperture areas.  However, all of them work in the same direction.  It is therefore not surprising that DESI should be more efficient than prior surveys, but it was not expected that the difference in redshift measurement efficiency would be so high.

\subsection{Implications for Spectroscopy in Support of Photometric Redshifts}\label{sec:updated-times}

Previous work by \citet{Newman2015SpectroscopicNeeds} and \citet{2022BlazekSnowmass} estimated the total telescope time that would be required to obtain a fiducial magnitude-limited spectroscopic sample designed to enable accurate \pz training and calibration using a variety of existing and proposed facilities. However, those estimates were scaled from the performance of the Keck/DEIMOS spectrograph as used in the DEEP2 survey. However, we have found that the redshift measurement efficiency of DESI is significantly higher than what would be expected from this scaling; accordingly, estimates of the required time should be reassessed for any instrument whose performance would be similar to DESI's.  

In this section we therefore provide time estimates for the fiducial survey from \citet{2022BlazekSnowmass} which are scaled from the measured performance of DESI rather than Keck/DEIMOS, which should be more appropriate for any instrument and observing strategy which more closely matches the DESI-Deep pilot program.  We closely adhere to all other assumptions made in that work. We also add survey time estimates for several new facilities and have updated instrument specifications to reflect their current designs.  

Following \citet{2022BlazekSnowmass}, we calculate the amount of time required for a benchmark survey designed to obtain spectroscopic samples for \pz training and calibration for the 10-year LSST sample. The survey goals are to acquire spectra for at least 20,000 objects down to a magnitude of $m_{i} = 25.3$ with a $\sim 90\%$ redshift measurement success rate. To mitigate the effects of cosmic variance, these targets must be distributed across either 15 widely separated fields for instruments with a field-of-view (FoV) $\leq 1 \deg^2$, 6 fields for $1\deg^2<\mathrm{FoV}<3\deg^2$, or 4 fields for instruments with FoV$\geq3 \deg^2$. At a minimum, each field is required to be at least $0.1\deg^2$ in area, which translates to a minimum total sky area of 1.5 $\deg^2$ for such a survey.

To calculate the total survey duration for different instruments, we first determine the required per-object exposure time with DESI ($\widetilde{T}_{\mathrm{exp}}$) using Equation~\ref{eq:desi-exp}. We scale the time required for DESI by the ratio of the effective light collecting areas of the Mayall telescope and the telescope hosting a given instrument. For any multiplexed observation, we assume that the exposure time required is dictated by the faintest objects in the fiducial survey (i.e., that faint objects are common enough that they will be included within any pointing/instrument configuration; since galaxy number counts rise geometrically with magnitude, this should be the case). The total survey duration is then calculated by considering the time required to meet a given requirement of the fiducial survey -- the number of fields required (the ``Field-limited'' case), the total survey area (the ``FoV-limited'' scenario), or the total number of objects with spectra (the ``multplex-limited'' case).  The survey time required to meet all requirements will then be the maximum of the time required for each of these scenarios.  

\textit{Field-Limited Case}: For an instrument with a large FoV and high multiplexing, the total observing time needed, $T_{\mathrm{obs}}$, will be determined by the number of fields, $N_{\mathrm{fields}}$, required to mitigate sample/cosmic variance:
\begin{eqnarray}
    T_{\mathrm{obs}} = N_{\mathrm{fields}} 
    &\times& \widetilde{T}_{\mathrm{exp}}(m_{i},p_{\rm success}) \label{eq:t_field} \\ 
    &\times& \frac{\mathrm{Collecting~Area_{DESI}}}{\mathrm{Collecting~Area_{Instrument}}} \nonumber
\end{eqnarray}
where $\mathrm{Collecting~Area_{DESI}}$ is set to 8.66~m$^2$. We take $N_{\mathrm{fields}}$ to be 15, 6, or 4, depending on the instrument's FoV, as described above.

\textit{FoV-Limited Case.} For an instrument with high multiplexing but a small FoV, the total observing time may be determined by the number of pointings needed to cover the required sky area:
\begin{eqnarray}
    T_{\mathrm{obs}} = \frac{\mathrm{Total~Survey~Area}}{\mathrm{FoV~Area}} 
    &\times& \widetilde{T}_{\mathrm{exp}}(m_{i},p_{\rm success}) \label{eq:t_fov} \\ 
    &\times& \frac{\mathrm{Collecting~Area_{DESI}}}{\mathrm{Collecting~Area_{Instrument}}} \nonumber
\end{eqnarray}
where the Total Survey Area is set to 1.5~deg$^2$.

\textit{Multiplex-Limited Case.} For an instrument with a large FoV but low multiplexing, the total observing time will be determined by the number of pointings needed to observe the total number of targets in the fiducial sample:
\begin{eqnarray}
    T_{\mathrm{obs}} = \frac{\mathrm{Number~of~Objects}}{\mathrm{Multiplexing}} 
    &\times& \widetilde{T}_{\mathrm{exp}}(m_{i},p_{\rm success}) \label{eq:t_mux} \\ 
    &\times& \frac{\mathrm{Collecting~Area_{DESI}}}{\mathrm{Collecting~Area_{Instrument}}} \nonumber
\end{eqnarray}
where the total Number of Objects is set to 20,000.

In Table~\ref{table:photoz_times}, we provide the final time estimates for a list of telescope and spectroscopic instrument pairs. This represents an update of Table 2 of \citet{2022BlazekSnowmass}.  All estimates are given in units of ``dark years'', where one dark year is defined as 365 5.33-hour dark-time nights (assuming an average of eight hours per night with one-third of the time lost due to weather and instrumental overheads). Dark-time is generally defined as the time after sunset when the sun is significantly below ($\gtrsim15\deg$) the horizon, the moon phase is not near full, and the moon is significantly far away ($\gtrsim50\deg$) from the field being observed (cf.~\citealt{Schlafly2023SurveyOps} for DESI's definition of dark-time). 

From Table~\ref{table:photoz_times}, we find that the required number of dark years is significantly ($\sim3\times$) reduced for each facility compared to the estimates in \citet{2022BlazekSnowmass} when assuming DESI-like efficiency. Given the ongoing progress in technology, it could be achieved with sufficient instrumental stability and data processing quality equivalent to what has been implemented with DESI. It is therefore reasonable to assume that future spectroscopic facilities will be at least as efficient, if not more. This provides reassurance that spectroscopic samples for \pz training and calibration of faint galaxy populations can be obtained in a feasible amount of time. We also note that DESI itself will be one of the most efficient instruments for obtaining such a sample.

\begin{deluxetable*}{ccccc}[!h]

\tabletypesize{\footnotesize}
\tablecaption{Total effective exposure time required to obtain spectroscopic data sets for \pz training and calibration using various spectroscopic facilities, assuming a DESI-like redshift measurement efficiency\label{table:photoz_times}.}

\tablehead{ \colhead{Telescope/Instrument} & \colhead{Collecting Area}  & \colhead{FoV Area}  & \colhead{Multiplex} & \colhead{Total time} \\
 & \colhead{[m$^2$]} & \colhead{[deg$^2$]} &  & \colhead{(Dark Years)}\\}
 \startdata
    Mayall / DESI & 8.66 & 8 & 5,000 & 0.55 \\ 
    VISTA / 4MOST & 11.99 & 4.2 & 1,624 & 1.21 \\
    WHT / WEAVE & 12.7 & 3.1 & 960 & 1.94 \\
    6m-class / Spec-S5* & 19 & 3.7 & 12850 & 0.12\tablenotemark{a} \\ 
    6.5m-class / MegaMapper* & 28.0 & 7.1 & 26,100 & 0.17 \\ 
    6.5m-class / MUST* & 30 & 6.16 & 21170 & 0.16 \\
    VLT / MOONS* & 52 & 0.55 & 1000 & 0.45 \\ 
    Subaru / PFS & 52.8 & 1.25 & 2,400 & 0.19 \\ 
    Keck / DEIMOS & 72.0 & 0.018 & 150 & 6.8\tablenotemark{b} \\ 
    Keck / FOBOS* & 72.0 & 0.1 & 1,800 & 0.25 \\ 
    12m-class / WST* & 113 & 3.1 & 20000 & 0.04 \\ 
    GMT / MANIFEST + GMACS* & 368 & 0.087 & 380 & 0.17 \\ 
    TMT / WFOS* & 633.45 & 0.007 & 100 & 0.4 \\ 
    ELT / MOSAIC Optical* & 978.0 & 0.01 & 200 & 0.18\tablenotemark{c} \\ 
    ELT / MOSAIC NIR* & 978.0 & 0.01 & 100 & 0.24 \\ 
\enddata
\tablenotetext{a}{Spec-S5 is proposed to be two identical telescopes and spectroscopic instruments. The quoted time assumes that both facilities are observing in parallel. The telescope parameters quoted are for each individual observatory.}
\tablenotetext{b}{The estimate for time required with Keck/DEIMOS has not been updated to reflect DESI-like efficiency as the original numbers from \citet{2022BlazekSnowmass} were calculated based on Keck/DEIMOS data.}
\tablenotetext{c}{For ELT, observations are required with both the optical and near-IR settings to achieve the required wavelength coverage, increasing total time required.}
\tablenotetext{*}{The telescope/instrument has been proposed but is not operational as of writing this document (i.e., early 2026).}
\tablecomments{The acronyms used in the table are defined as follows, DESI: Dark Energy Spectroscopic Instrument, VISTA: Visible and Infrared Survey Telescope for Astronomy, 4MOST: 4-metre Multi-Object Spectroscopic Telescope, WHT: William Herschel Telescope, WEAVE: WHT Enhanced Area Velocity Explorer, Spec-S5: The Stage-5 Spectroscopic Experiment, MUST: Multiplexed Survey Telescope, VLT: Very Large Telescope, MOONS: Multi-Object Optical and Near-IR Spectrograph, PFS: Prime Focus Spectrograph, DEIMOS: DEep Imaging Multi-Object Spectrograph, FOBOS: Fiber-Optic Broadband Optical Spectrograph, WST: Widefield Spectroscopic Telescope, GMT: Giant Magellan Telescope, MANIFEST: Many Instrument Fiber System, GMACS: GMT Multi-object Astronomical and Cosmological Spectrograph, TMT: Thirty Meter Telescope, WFOS: Wide-Field Optical Spectrometer, ELT: Extremely Large Telescope, MOSAIC: Multi-Object Spectrograph for Astrophysics, Intergalactic medium studies and Cosmology, NIR: Near Infrared.
}
\tablerefs{DESI:\citet{DESI2022Overview}, 4MOST:\citet{deJong20194MOST}, WEAVE:\citet{Jin2024WEAVE}, Spec-S5:\citet{Besuner2025Spec-S5}, MegaMapper:\citet{Schlegel2022Megamapper}, MUST:\citet{Zhao2024MUST}, MOONS: \citet{Cirasuolo2020Moons}, PFS:\citet{Tamura2016PFS}, DEIMOS: \citet{Faber2003Deimos}, FOBOS:\citet{Bundy2019Fobos}, WST:\citet{Bacon2024WST}, MANIFEST:\citet{Zafar2022MANIFEST}, GMACS: \citet{Depoy2018GMACS}, WFOS:\citet{Steidel2022WFOS}, MOSAIC:\citet{Roser2024MOSAIC}.}

\end{deluxetable*}

\subsection{Design of a Potential \pz Survey with DESI (DESI-Deep)}\label{sec:future-survey}

The efficiency of DESI for measuring redshifts, as demonstrated in this work, can enable a dedicated survey of faint galaxies that would provide a transformative data set to the community with a modest investment of DESI observing time. The resulting data set would enhance the cosmological constraints of major imaging surveys and benefit nearly all areas of extragalactic science by characterizing the spectral energy distributions of a large, magnitude-limited sample of galaxies up to high redshifts. This data set could be used not only to train and calibrate \pzs but also to develop similar methods for determining galaxies' physical properties from photometric information (e.g., training algorithms for predicting stellar masses such as that presented in ~\citealt{Zhou2023LRGTargetSelection}). 

In this subsection, we describe an example design for a dedicated survey program with DESI, which we call the ``DESI-Deep Survey'' hereafter, which provides an example of what sort of training sets the instrument would provide.  This design builds on the pilot program used in this work and addresses the critical lack of representative spectroscopic samples reaching depths comparable to lensing samples from early LSST data (let alone the full 10-year survey depth).  

An ideal DESI-Deep survey would target a magnitude-limited sample of galaxies with $18 \leq m_{i} \leq 24.5$ and employ minimal selection cuts to ensure a representative sample. Targets can be selected from LSST photometry when available or from existing wide-area surveys like the HSC-SSP to enable immediate implementation; analysis of the sample would require LSST-based color information (e.g., for creating tomographic redshift bins or forward modeling any incompleteness) but targeting would not. The magnitude-limited sample would be downsampled to have a more uniform distribution in $i$-band magnitude to ensure that targets are not dominated by the faintest objects. In order to mitigate the effects of sample/cosmic variance, the proposed survey would perform observations on two $\sim10$ square degree fields with a total exposure time of 25 hours per field, observing at least 1000 objects per unit magnitude and a total of 10,000 objects per field. The first phase of the survey is proposed to be conducted in two fields for easy logistics and running it in parallel to the main DESI survey but could be expanded to four or more fields in a future phase.

The target of observing 20,000 objects in total (10,000 per field) is motivated by the requirements for calibrating the redshift distributions for tomographic redshift bins used in cosmological analyses. Assuming a typical choice of five tomographic bins, this total sample size provides approximately 4,000 objects per bin. The primary goal is to precisely characterize the mean redshift of each tomographic bin, $\langle z \rangle$, as well as the spread in redshifts, $(\langle(z - \langle z \rangle)^2\rangle)^{1/2}$. Under the assumption that the observed sample is a random subsample of the underlying true distribution, the standard error on the population mean scales as $\sigma_{n(z)} / \sqrt{N}$, where $\sigma_{n(z)}$ is the standard deviation of the redshift distribution within the bin (where typically $\sigma_{n(z)} \approx 0.1$ for a redshift bin) and $N$ is the number of objects. Similarly, the standard error on the standard deviation scales as $\sigma_{n(z)} / \sqrt{2N}$, so it is always smaller (but has similar calibration requirements to the mean, as found in \citealt{DESCSRD}).  A sample of 4,000 objects per bin with a success rate of 75\% would thus reduce the standard error of $\langle z \rangle$ to $\sim 2 \times 10^{-3}$ and the standard error of $(\langle(z - \langle z \rangle)^2\rangle)^{1/2}$ to $\sim 1.4 \times 10^{-3}$, meeting the requirements for stage IV dark energy experiments \citep{DESCSRD}.

Photometric redshift training and calibration are most effectively done in fields with deep photometry and rich multi-wavelength imaging, which enables the breaking of redshift degeneracies and the characterization of the impact of photometric errors \citep{Wright2020SOMDirectCal,Myles2021DESY3Photoz}. The ideal fields for a DESI-Deep program are therefore the two equatorial Rubin Observatory LSST Deep Drilling Fields (DDFs) --- COSMOS and XMM-LSS, as they will have exceptionally deep photometry from both Rubin and the Roman Space Telescope and extensive ancillary data. Of the LSST DDFs, the only other field accessible at all to DESI is ECDF-S, but it could only be observed at high air masses, greatly reducing observational efficiency.

A DESI-Deep survey can employ an observing strategy that is very similar to that used for the pilot sample. Effective exposure times of 1000 seconds per observation, as are used for the main DESI survey, should still be a good choice for such a program. The full field area would be tiled with dithered pointings in a rosette-like pattern in order to ensure that each target is observed with different fibers in different exposures.  This serves to suppress any instrumental systematics that vary from fiber to fiber while maximizing both the number of unique targets observed and the cumulative exposure time for high-priority targets. The observations could be spread over multiple months or years and potentially be paired with other programs --- such as instrument calibration or transient host redshift measurement --- that require multiple visits to the same patch of sky over a long period of time. 

\begin{figure}
    \centering
    \includegraphics{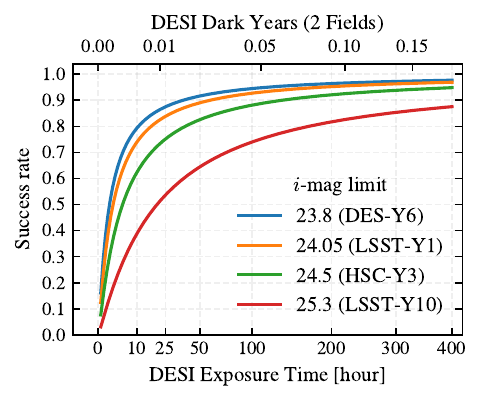}
    \caption{Predicted redshift measurement success rates plotted as a function of the total amount of dark time required for various magnitude limits. The magnitude limits are chosen to roughly represent the magnitude limits of the samples of various frontier weak lensing experiments. The lower horizontal axis displays the exposure times needed per object, while the upper horizontal axis shows the total time required to collect a magnitude-limited sample in the field-limited regime for 2 fields, measured in units of dark-years (8 hours of observations per night for 365 days of which 1/3rd is lost due to weather and instrument overheads. The extrapolation of the success rate beyond $\sim 7$ hours assumes the SNR continues to follow expectations for background-dominated observations and is not affected by instrumental systematics. }
    \label{fig:model-success-time}
\end{figure}

The faint magnitude limit for the proposed DESI-Deep survey was chosen to balance two goals: creating a dataset that is immediately useful for near-term science while ensuring that the exposure time per object needed to obtain a high redshift success rate is not too long. Based on our predictive model (Equation~\ref{eq:pr-success-fit}), we expect a DESI-Deep survey with 25-hour maximum exposure times per target to achieve an overall success rate of at least 75\% for objects with $m_{i}=24.5$ (which is the same as the lensing sample depth for the HSC Year 3 data set). We anticipate this will exceed 80\% at the anticipated LSST Year 1 lensing sample depth ($m_{i} = 24.05$) and be 90\% or higher for targets with $i$-band magnitudes below 23. In Figure~\ref{fig:model-success-time} we show the expected success rates as a function of exposure time required for a variety of different magnitude limits. The magnitude limits are chosen to roughly represent expected source samples from various frontier weak lensing experiments.

Any spectroscopic survey will have redshift measurement failures, even with very long exposure times. Moreover, these failures do not occur randomly but tend to affect galaxies with specific colors or at particular redshifts, which can introduce biases in the inferred redshift distributions if not accounted for. We note that galaxies for which a reliable redshift measurement could not be obtained with DESI could be followed up in the future with larger telescopes with infrared spectrographs to determine the nature of redshift failures and further increase the value of a DESI-Deep data set. 

\subsection{Impact of Incompleteness of Spectroscopic Samples on Cosmology }

While a complete quantification of the complex and non-linear effects of redshift failures is beyond the scope of this work, it is valuable to assess their potential impact on cosmological parameter constraints. To this end, we here use a cosmological forecasting framework to assess the impact of such failures for DESI-Deep-like surveys under two extreme scenarios that represent best- and worst-cases for occurrences in redshift failures.  

Specifically, we use the Fisher forecasting framework developed in \citet{Zhang2025forecast} to predict the Dark Energy Task Force Figure of Merit (FoM;\citealt{DETF2006}) for LSST Year-1 and Year 10 3$\times$2pt analyses (i.e., the combination of weak lensing, galaxy-galaxy lensing, and galaxy clustering measurements) in the $w_0$–$w_a$CDM model. We calculate the FoM as the inverse of the square root of the determinant of the covariance matrix for the $w_0$ and $w_a$ parameters to be inferred in a cosmological analysis. It is therefore proportional to the Dark Energy Task Force Figure of Merit defined as the reciprocal of the area of the error ellipse enclosing the 95\% confidence limit in the $w_0$–$w_a$ plane. A higher value of FoM corresponds to a lower area for the error ellipse and corresponds to a cosmological experiment with greater constraining power.  Our forecasts use a fiducial cosmology based on Planck 2018 $\Lambda$CDM parameters \citep{Planck2018Cosmology}, and the baseline source redshift distribution, $n(z)$, excludes catastrophic outliers, following \citet{Zhang2025forecast} and \citet{DESCSRD}.

The forecasts are computed for various redshift failure scenarios by modifying the effective number density, $n_{\rm eff}$, of source galaxies in the five tomographic bins assumed to be used for the 3$\times$2pt analysis. We assume that the reduction of $n_{\rm eff}$ is entirely due to the impact of redshift measurement failures, so that the value of $n_{\rm eff}$ relative to the ideal case corresponds to the success rate of the redshift measurements. This corresponds to the assumption that we can (e.g., based on colors) separate out the population of galaxies which has complete redshift characterization from spectroscopy, and that lensing and clustering analyses would only use those objects; thus, a survey with higher redshift measurement success rates would enable lensing analyses that utilize more objects. Although this constitutes a strong assumption, the availability of high SNR photometry across approximately six bands, as will be provided by LSST, should render such characterization feasible.

Specifically, we scale the covariance matrix for LSST 3$\times$2pt analyses assuming that shape noise dominates the lensing contribution; in that case, the noise per shear component scales as $1/\sqrt{n_{\rm eff}}$. We assume five tomographic source bins divided such that each contains the same number of (pre-failure) galaxies, and redshift failures in one bin are treated as independent of the others. The priors on redshift bias and scatter are fixed across all scenarios, derived from $N_{\rm samples} = 5000$ reference redshift samples (see Section 3.5 of \citealt{Zhang2025forecast} for details). The forecasts include a redshift-dependent intrinsic alignment model and marginalize over intrinsic alignment, photometric redshift, and galaxy bias parameters when computing the FoM on cosmological parameters.

We forecast the relative FoM under two extreme scenarios; realistic observations should yield results somewhere between these possibilities. In the first, redshift failures occur evenly across all tomographic bins, resulting in a uniform decrease in $n_{i,\rm eff}$ for all bins. This represents a best-case scenario in which redshift failures occur randomly and equally at all redshifts. In the second case, all failures are confined to the highest redshift bins; e.g., we assume that all redshift failures happen either in the last redshift bin; or, if the failure rate exceeds the number of objects in that bin, that they occur  in the last two bins, or the last three bins if the number is even greater; etc.  For failure rates that are not a multiple of 20\%, we interpolate between the FoM values for 0\%, 20\%, 40\%, etc. failure rates (i.e., the cases where a bin is entirely lost); this approximates the results from dropping the highest-redshift objects from analysis entirely, rather than still using a subset of objects at very high $z$.
This scenario represents the other extreme where redshift failures are all concentrated at the highest redshifts, potentially causing entire redshift ranges to be impossible to utilize in analyses.  In all cases, we define the relative FoM as the ratio between the FoM when $n_{i,\rm eff}$ is reduced to the FoM when there has been no reduction.

\begin{figure*}[!h]
    \centering
    \includegraphics[]{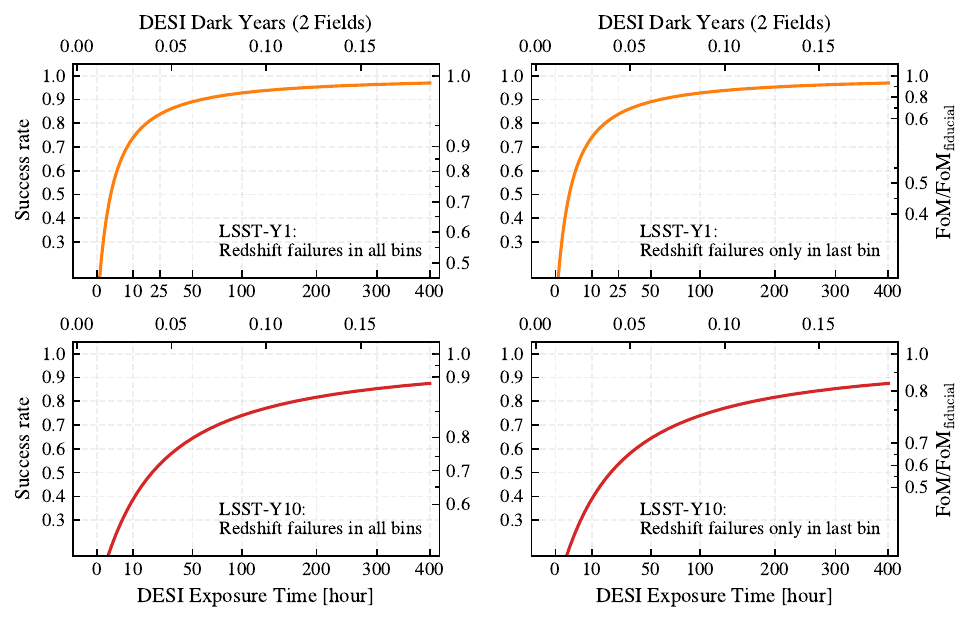}
    \caption{The relationship between exposure time with the DESI instrument, redshift measurement success rate, and the LSST survey cosmological constraining power as quantified by the relative DETF figure of merit (FoM/FoM$_\mathrm{fiducial}$), degraded due to parts of parameter space lacking calibration redshifts due to spectroscopic measurement failures. The top panels correspond to an LSST Year-1-like 3$\times$2pt (weak lensing plus galaxy clustering) analysis, while the bottom panels correspond to an LSST Year-10-like analysis. The left panels show the predicted redshift success rate (left-vertical axis) or relative FoM (right-vertical axis) as a function of DESI exposure time (lower-horizontal axis) for the case where redshift failures occur randomly at all redshifts, while the right panels display the same but for the case where the redshift failures only occur at the very highest redshifts. The left panels correspond to the best case scenario for a spectroscopic survey while the right panels correspond to a worst case scenario; real surveys will fall somewhere in between these limits. We have added the total time required for a (field-limited) spectroscopic survey on the upper-horizontal axis of each of these plots for easy comparison to the discussions in Section~\ref{sec:updated-times}. This figure directly connects the investment in spectroscopic observing time to its impact on cosmology, providing a tool to assess the trade-offs in designing a spectroscopic survey to obtain data for \pz training and calibration. For an LSST Year-1 like analysis, achieving an 80\% spectroscopic success rate as is the goal for the DESI-Deep survey (Section \ref{sec:future-survey}), results in a relative FoM value between 0.96 (for uniform failures) and 0.64 (for worst-case failures). Assuming a roughly circular error ellipse in the $w_0$-$w_a$ plane, this FoM degradation corresponds to an increase in the size of cosmological parameter error bars of 2\% to 25\% compared to an ideal survey with no redshift failures. This level of statistical uncertainty is well within the error budget for current and near-term weak lensing experiments.}
    \label{fig:time-cosmology}
\end{figure*}

The results of this forecasting exercise are presented in Figure~\ref{fig:time-cosmology}. This figure shows the relative FoM as a function of the overall spectroscopic success rate. We show the FoM degradation for the best-case and worst-case failure scenarios, evaluated for both the LSST Year-1 and Year-10 survey depths. Using the predictive model developed in this work, we can directly link the success rate to a required survey exposure time and thereby connect the observing time invested in a \pz survey directly to its impact on cosmological constraining power. While this plot informs the design of our proposed DESI-Deep survey, it also serves as a general tool for forecasting the cosmological impact of redshift failures for other future spectroscopic \pz\ training surveys.

We note that for an LSST Year-1 like analysis, achieving an 80\% spectroscopic success rate as is the goal for the DESI-Deep survey (Section \ref{sec:future-survey}) results in a relative FoM value between 0.96 (for uniform failures) and 0.64 (for worst-case failures). Assuming a roughly circular error ellipse in the $w_0$-$w_a$ plane, this FoM degradation corresponds to an increase in the size of cosmological parameter error bars of 2\% to 25\% compared to an ideal survey with no redshift failures. This level of statistical uncertainty is well within the error budget for current and near-term weak lensing experiments, demonstrating that a spectroscopic data set from the proposed DESI-Deep survey provides a powerful and feasible path toward robust cosmological constraints.  Future surveys with longer exposure times and correspondingly higher completeness rates can further reduce the impact of failures.

\section{Summary}\label{sec:summary}

In this work, we have presented the results from the DESI-Deep pilot program, which was designed to obtain deep spectroscopy for a faint, magnitude-limited sample of galaxies ($22 \leq m_i \leq 24.5$). Our goal was to assess DESI's redshift measurement performance for such samples and its potential to provide the crucial, representative spectroscopic samples needed for photometric redshift training and calibration for next-generation imaging surveys such as LSST. We find that DESI is remarkably efficient at this task; with just two hours of exposure time, it can match or exceed the redshift success rates of the DEEP2 and DEEP3 surveys conducted with the DEIMOS spectrograph on the Keck 10m telescope. Furthermore, we demonstrated that the signal-to-noise ratio of the spectra scales as expected for background-limited observations even for our longest exposures ($\sim 7$ hours) and faintest targets. This indicates that DESI has not yet hit a systematic floor and that even deeper observations are feasible. We attribute this high performance to a combination of DESI's excellent throughput, superior astrometry, advanced data processing algorithms, and a dithering strategy that minimizes instrumental systematics.

Building on these results, we have developed a quantitative model that predicts the redshift success rate as a function of galaxy magnitude and exposure time. Using this model, we have updated estimates of the time required to obtain a benchmark spectroscopic sample for the full 10-year LSST dataset, finding that a Stage V spectroscopic facility with DESI-like performance but a larger mirror (e.g., Spec-S5) could achieve this in as few as 50 dark nights. We propose a dedicated ``DESI-Deep Survey'' to obtain a more limited version of such a sample, targeting $\sim$20,000 galaxies down to $m_{i}=24.5$ in two LSST Deep Drilling Fields. A cosmological forecasting analysis shows that the resulting dataset, even with realistic redshift failure rates, would be transformative for cosmology, dramatically improving the precision of cosmological constraints from LSST and providing a lasting legacy resource for all of extragalactic astronomy.

\section*{DATA AVAILABILITY}

Code and data used in this work are available at: \url{https://biprateep.github.io/desi-deep-pilot/}. At present, only the targeting, redshift, and visual-inspection catalogs, along with spectral-derived data products, are being made available. The raw spectra will be distributed as part of the DESI Data Release 2. The redshift catalog for this data set is also included in the compilation of spectroscopic redshifts in the DESI-COSMOS and DESI-XMMLSS fields presented by \citet{Ratajczak2026DESI}.

\begin{acknowledgments}
B. Dey is a postdoctoral fellow at the University of Toronto in the Eric and Wendy Schmidt AI in Science Postdoctoral Fellowship Program, a program of Schmidt Sciences.  B. Dey and B. H. Andrews acknowledge the support of the National Science Foundation under Grant No.\ AST-2009251.  Any opinions, findings, and conclusions or recommendations expressed in this material are those of the author(s) and do not necessarily reflect the views of the National Science Foundation. The efforts of J. A. Newman were supported by grant DE-SC0007914 from the U.S. Department of Energy Office of Science, Office of High Energy Physics.

This material is based upon work supported by the U.S. Department of Energy (DOE), Office of Science, Office of High-Energy Physics, under Contract No. DE–AC02–05CH11231, and by the National Energy Research Scientific Computing Center, a DOE Office of Science User Facility under the same contract. Additional support for DESI was provided by the U.S. National Science Foundation (NSF), Division of Astronomical Sciences under Contract No. AST-0950945 to the NSF’s National Optical-Infrared Astronomy Research Laboratory; the Science and Technology Facilities Council of the United Kingdom; the Gordon and Betty Moore Foundation; the Heising-Simons Foundation; the French Alternative Energies and Atomic Energy Commission (CEA); the National Council of Humanities, Science and Technology of Mexico (CONAHCYT); the Ministry of Science, Innovation and Universities of Spain (MICIU/AEI/10.13039/501100011033), and by the DESI Member Institutions: \url{https://www.desi.lbl.gov/collaborating-institutions}. Any opinions, findings, and conclusions or recommendations expressed in this material are those of the author(s) and do not necessarily reflect the views of the U. S. National Science Foundation, the U. S. Department of Energy, or any of the listed funding agencies.

The authors are honored to be permitted to conduct scientific research on I'oligam Du'ag (Kitt Peak), a mountain with particular significance to the Tohono O’odham Nation.

The Hyper Suprime-Cam (HSC) collaboration includes the astronomical communities of Japan and Taiwan, and Princeton University. The HSC instrumentation and software were developed by the National Astronomical Observatory of Japan (NAOJ), the Kavli Institute for the Physics and Mathematics of the Universe (Kavli IPMU), the University of Tokyo, the High Energy Accelerator Research Organization (KEK), the Academia Sinica Institute for Astronomy and Astrophysics in Taiwan (ASIAA), and Princeton University. Funding was contributed by the FIRST program from the Japanese Cabinet Office, the Ministry of Education, Culture, Sports, Science and Technology (MEXT), the Japan Society for the Promotion of Science (JSPS), Japan Science and Technology Agency (JST), the Toray Science Foundation, NAOJ, Kavli IPMU, KEK, ASIAA, and Princeton University. 

This paper makes use of software developed for Vera C. Rubin Observatory. We thank the Rubin Observatory for making their code available as free software at \url{http://pipelines.lsst.io/}.

This paper is based on data collected at the Subaru Telescope and retrieved from the HSC data archive system, which is operated by the Subaru Telescope and Astronomy Data Center (ADC) at NAOJ. Data analysis was in part carried out with the cooperation of Center for Computational Astrophysics (CfCA), NAOJ. We are honored and grateful for the opportunity of observing the Universe from Maunakea, which has the cultural, historical and natural significance in Hawaii.

This research was supported by the Munich Institute for Astro-, Particle and BioPhysics (MIAPbP) which is funded by the Deutsche Forschungsgemeinschaft (DFG, German Research Foundation) under Germany's Excellence Strategy – EXC-2094 – 390783311.

The authors advocate for the judicious and ethical use of artificial intelligence (AI) based tools in scientific discourse. In accordance with the \href{https://aas.org/posts/news/2023/03/use-chatbots-writing-scientific-manuscripts}{editorial guidelines} of the American Astronomical Society on the appropriate use of AI-based writing tools, the authors confirm that AI assistance (with Google Gemini; \citealt{GoogleGemini2023}) was utilized in the preparation of this manuscript solely to polish, condense, and edit our original research. The authors retain the full responsibility for the accuracy, integrity, and content of this work.

\end{acknowledgments}





%
\facilities{Dark Energy Spectroscopic Instrument (DESI;\citealt{DESI2022Overview}), Hyper Suprime-Cam (HSC; \citealt{HSC2018HSCInstrument}), DEIMOS \citep{Faber2003Deimos}}.

\software{Astropy\citep{AstropyEtal2013, AstropyEtal2018,AstropyEtal2022}, Google Gemini \citep{GoogleGemini2023}, Matplotlib \citep{Hunter2007Matplotlib}, Numpy \citep{CharlesEtal2020Numpy}, Pandas \citep{mckinney2010Pandas,reback2020Pandas}, Scikit-Learn \citep{PedregosaEtal2011Sklearn}, Scipy \citep{VirtanenEtal2020Scipy}, Statsmodels \citep{seabold2010statsmodels}.
          }

\clearpage
\appendix
\section{Additional Observations in the DESI-Hercules Field} \label{app:hercules}

As part of the DESI-Deep pilot program, we conducted additional observations in the DESI-Hercules field (defined in \autoref{sec:data}). Due to the low quality of the data obtained in this field, it was not used for the results in the main body of the article. In this appendix, we document this sample and present selected results which include this dataset along with the other two fields.

\begin{figure*}[!h]
    \centering
    \includegraphics{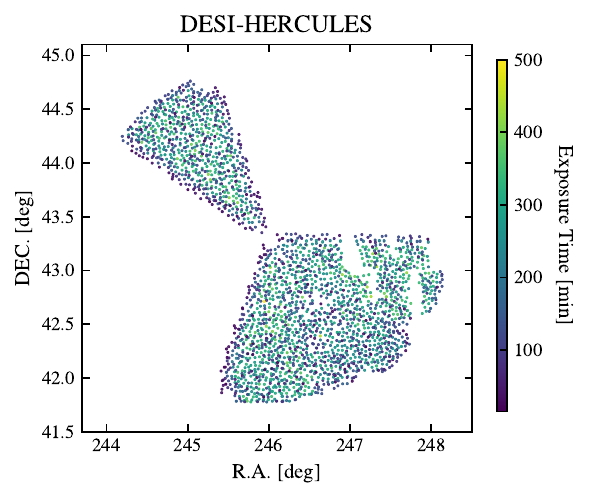}
    \caption{Locations of the 2511 objects observed in the DESI-Hercules ($244\degree \leq$R.A.$\leq248.5\degree$; $41.5\degree\leq$DEC.$\leq45\degree$) field. The colors indicate the total effective exposure time for the object. The empty region in the focal plane is due to certain spectrographs not being operational.}
    \label{fig:fibers_on_sky_hercules}
\end{figure*}

The targets in the DESI-Hercules field were selected based upon HSC photometry using the same criteria  as applied in the other regions, and all data processing and analysis procedures were identical to those used for our other results. Figure~\ref{fig:fibers_on_sky_hercules} shows the on-sky locations of the observed targets in all three fields, along with the effective exposure times per object. The empty regions in the focal plane correspond to spectrographs that were unavailable during these observations. The data collected came from spectrographs that had not been cooled to their optimal operating temperatures.

\begin{figure*} [!h]
    \centering
    \includegraphics{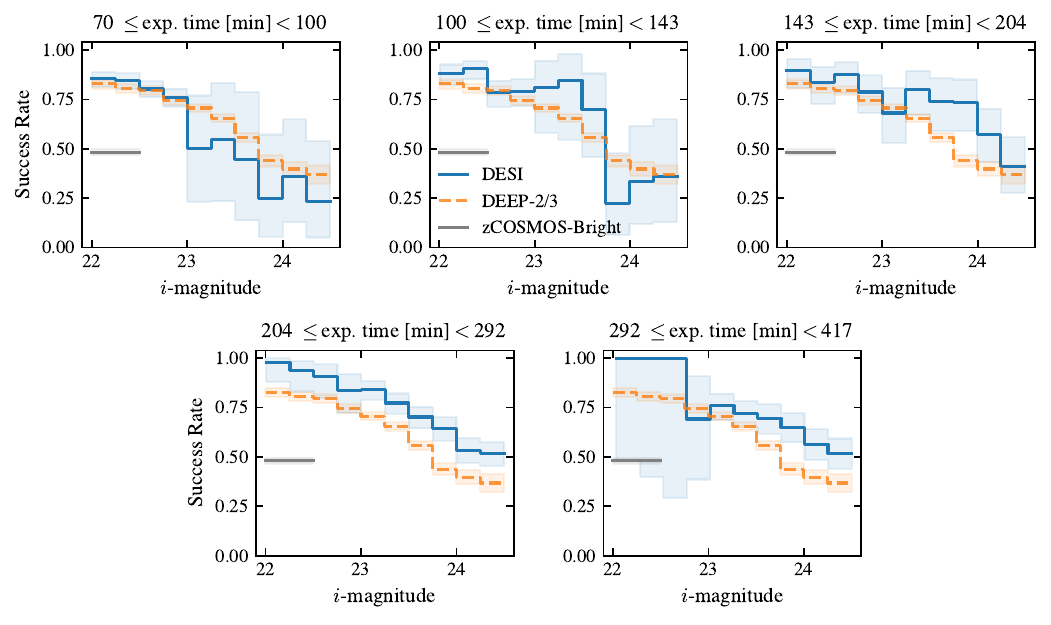}
    \caption{As Figure~\ref{fig:i-mag-success}, but with data from the DESI-Hercules field included.  The plot shows measurement success rate for the full DESI-Deep pilot sample as a function of $i$-band magnitude, with separate panels corresponding to different ranges of effective exposure time. The blue curve shows the success rate obtained from our observations, with the 95\% confidence interval shown using the shaded blue region. The success rates as a function of $i$-magnitude for the combined DEEP2 and DEEP3 survey datasets are plotted in orange for comparison, with the shaded orange area showing the corresponding 95\% confidence interval. The gray line denotes the average success rate for the zCOSMOS-Bright sample for objects with $i$-band magnitudes between 22 and 22.5. Compared to Figure \ref{fig:i-mag-success}, the inclusion of DESI-HERCULES data lowers the standard errors on the success rate measurements, but results in slightly lower success rates. However, at exposure times of about 2 hours, DESI’s success rate still matches that of the DEEP2 and DEEP3 galaxy redshift surveys.}
    \label{fig:i-mag-success-hercules}
\end{figure*}

\begin{figure*} [!h]
    \centering
    \includegraphics{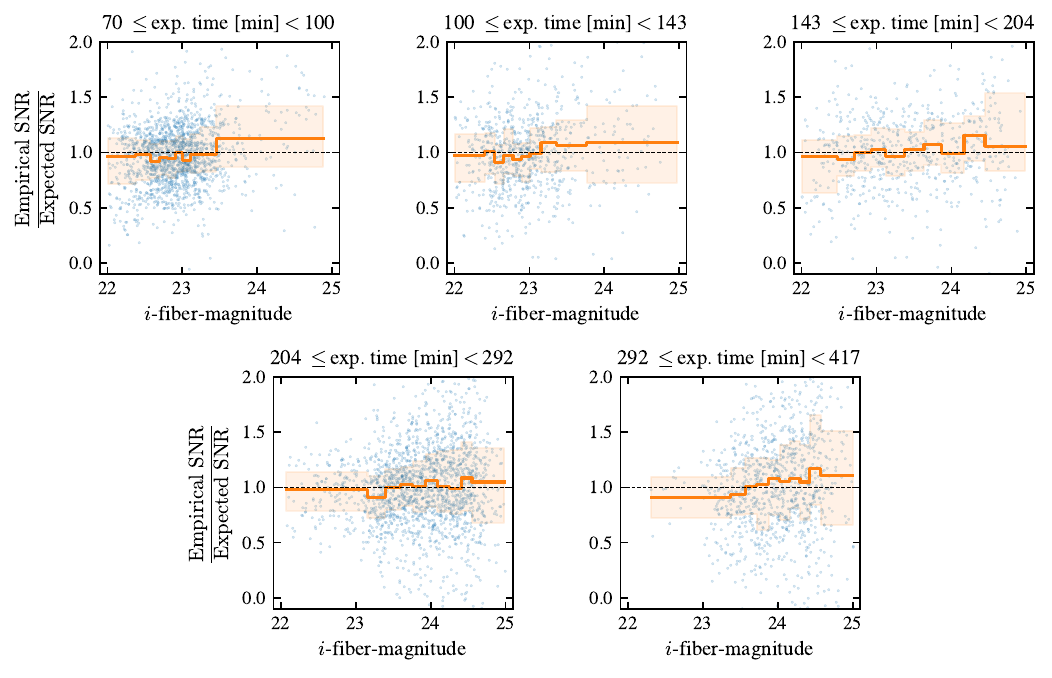}
    \caption{As Figure \ref{fig:snr-v-mag}, but with data from the DESI-Hercules field included. The ratio of empirically measured SNR and what we would expect from a background limited regime plotted as a function of $i$-fiber-magnitude in bins of exposure time for the full DESI-Deep pilot sample. Each blue dot represents a single object and the orange curve denotes the average in ten equal population bins. The shaded orange region shows the 95\% confidence interval on the mean. Compared to Figure \ref{fig:snr-v-mag} we observe a slightly worse scaling of the SNR especially at the fainter magnitudes. However, this is expected, given the non-optimal status of the instrument during the DESI-HERCULES observations.}
    \label{fig:snr-v-mag-hercules}
\end{figure*}

Figures~\ref{fig:i-mag-success-hercules} and \ref{fig:snr-v-mag-hercules} show DESI's redshift measurement success rate and the ratio of empirical to expected SNR, respectively, as a function of magnitude and exposure time, combining data from all three fields. Including these additional observations increases the total number of targets to approximately 7,370, which reduces the statistical errors on our binned measurements. However, the lower quality of this dataset also reduces the overall redshift measurement efficiency and weakens the SNR scaling due to the known issues with the Hercules data. Despite this, all main conclusions drawn in this work remain unchanged even after including this additional data.


\clearpage
\bibliography{astro_references,software_references,desi_references}{}
\bibliographystyle{aasjournalv7}


\global\suppressAffiliationsfalse
\allauthors
\end{document}